\documentclass[aps,prb,showpacs,floatfix,twocolumn,superscriptaddress]{revtex4-2}
\usepackage{graphicx}
\usepackage{dcolumn}
\usepackage{enumitem}
\usepackage{cancel}
\usepackage{bm}
\usepackage{tikz}
\usepackage{siunitx}
\usepackage{tablefootnote}
\usepackage{physics}
\usepackage{nicefrac}
\usepackage{booktabs}
\usepackage[%
  pdfauthor={Benedikt Placke},%
  pdfstartview=FitH,%
  breaklinks=true,%
  bookmarks=true,%
  colorlinks=true,%
  anchorcolor=black,%
  citecolor=red,
  filecolor=black,%
  menucolor=black,%
  urlcolor=blue,%
  linkcolor=red,%
 ]{hyperref}
\usepackage[english]{babel}
\usepackage{bm}
\usepackage{amssymb}
\usepackage[utf8]{inputenc}
\usepackage{newunicodechar}
\newunicodechar{，}{,}
\newunicodechar{−}{-}
\usepackage{color, soul}
\usepackage{amsmath}
\usepackage{amstext}
\usepackage{latexsym}
\usepackage{graphicx}
\usepackage{mathtools}
\usepackage[ruled]{algorithm2e}
\makeatother
\begin{document}
 \author{Changzhi Zhao}
\affiliation{College of Physics, Taiyuan University of Technology, Shanxi 030024, China}

\author{Wanzhou Zhang}
 \email{zhangwanzhou@tyut.edu.cn}
 \affiliation{College of Physics, Taiyuan University of Technology, Shanxi 030024, China}
 \author{Yuan Huang}
 \affiliation{DP Technology, Beijing 100080, China}
  \author{Chengxiang Ding} \email{dingcx@ahut.edu.cn}
  \affiliation{School of Science and Engineering of Mathematics and Physics, Anhui University of Technology, Maanshan, Anhui 243002, China}

\author{Youjin Deng}%
\email{yjdeng@ustc.edu.cn}
\affiliation{
Hefei National Laboratory for Physical Sciences at the Microscale and Department of Modern Physics, University of Science and Technology of China, Hefei 230026, China}
\affiliation{Hefei National Laboratory, University of Science and Technology of China, Hefei 230088, China}


\date{\today}
\title{Emergent  critical phases of the Ashkin–Teller model on the Union-Jack Lattice}

\begin{abstract}
The Ashkin-Teller (AT) model is a classic spin model in statistical mechanics. For traditional homogeneous lattices like triangular and kagome lattices, even when frustration exists, the model only has one ferromagnetic-paramagnetic critical line in the \(J>0\) and \(K<0\) region. However, in this paper,  for the Union Jack lattice, where the lattice coordination numbers are 4, 8, and 8 and which also contains a large number of small triangular units, using Metropolis Monte Carlo method, we find that, the critical line of the AT model splits into two Berezinskii–Kosterlitz–Thouless(BKT) boundaries, and a critical phase emerges in the intermediate region.  This phenomenon is the combined result of frustration, lattice inhomogeneity and the
two coupled spin degrees of freedom inherent to the AT
model. In detail, the novel critical phase characterized by a power-law decay of magnetization with system size, where the correlation length ratio $\xi/L$ remains finite even in the thermodynamic limit. We also introduce the susceptibility \(\widetilde{\chi} = \text{d}\langle m \rangle /\text{d}J\) as a key probe, and through this probe, pseudo-critical points \(J_c(L)\) are observed to scale proportionally to \((\ln L)^{-2}\), a behavior consistent with BKT criticality.
Since  superfluids, superconductors, and supersolids all possess quasi-long-range order and fall into the category of critical phases, our results could also inspire the exploration of such quantum phases.
\end{abstract}

\maketitle

\section{Introduction}
In this paper, we study the Ashkin-Teller (AT) model \cite{ashkin1943}, 
and the Hamiltonian is given by
\begin{equation}
\frac{\mathcal{H}}{k_BT} =-J\sum_{\langle i,j\rangle }(\sigma _{i} \sigma _{j}+\tau _{i} \tau _{j})-K\sum_{\langle i,j\rangle }\sigma _{i} \sigma _{j} \tau _{i} \tau _{j}.
\label{eq:ham}
\end{equation}
This is the reduced classical Hamiltonian~\cite{reduce} at finite temperature. 
We adopt natural units by setting $T=1$ and $k_B=1$ throughout the simulations.  This model can be viewed as the coupling of two identical lattices, where \( \sigma_i, \tau _i = \pm 1 \) are the two Ising spins, respectively. \( J \) denotes the two-spin coupling interaction, \( K \) represents the four-spin interaction, and \( \langle i,j \rangle \) indicates the nearest-neighbor interactions. Usually, a coupled spin variable \(s = \sigma \cdot \tau\) is also defined to characterize the phases.
The AT model has been experimentally implemented using Selenium adsorbed on Ni surface~\cite{ex1}， where the positions of the selenium atoms naturally form the required four-state configurations. Ashkin-Teller universality has been observed in multi-component Rydberg atom systems when the density wave phase with a period of four melts \cite{atom1, atom2}. Theoretical studies reveal fascinating critical phenomena in the AT model, including dynamics~\cite{36}, percolation transitions~\cite{percolation_at}, emergent symmetries~\cite{xy_3dat}, and connections to quantum states~\cite{zhu,quantum_at} and so on.

\begin{figure}[t]
    \centering
    \includegraphics[width=0.5\linewidth]{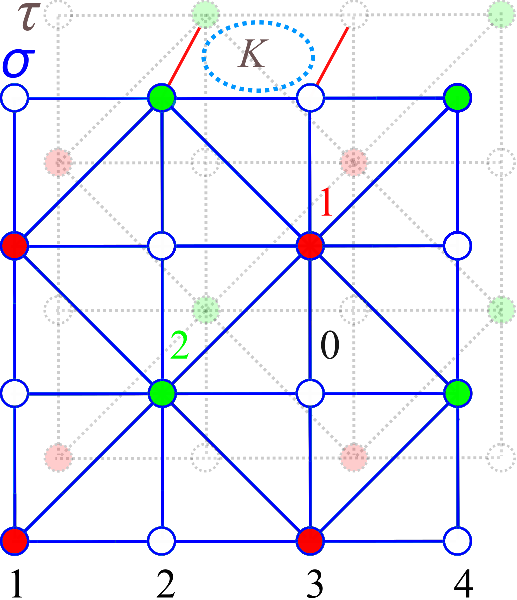}
 \caption{ Schematic of a 4$\times$4 double-layer UJ lattice with periodic boundary conditions. Circles of different colors denote sublattices 0, 1, and 2, respectively. They can be also labeled $A_4$, $B_8$ and $C_8$~\cite{XYmodel_UJ}. One layer corresponds to spins \( \sigma \), and the other to spins \(  \tau  \). The four-body interaction between layers is labeled by \( K \).}
    \label{fig:lattice}
\end{figure}
 Importantly, its phase diagrams exhibit exceptional richness~\cite{9,23,24} in diverse lattice geometries, as demonstrated in uniform systems such as triangular and kagome lattices~\cite{worm}, providing ideal platforms for exploring novel physics. In the square lattice, there are three phases: the ferromagnetic (FM) phase, the paramagnetic (PM) phase, and the phase where $\sigma$ and $\tau$ are paramagnetic while $s$ is in the antiferromagnetic (AFM) phase, in the parameter regimes  $J>0$ and $K<0$.
 Studying the phase diagrams of the AT model on inhomogeneous lattices is of particular interest, as such phase diagrams may have novel phases induced by structural inhomogeneity. 
 The UJ lattice~\cite{18}, is a paradigmatic non-uniform structure and thus an ideal platform to explore how structural inhomogeneity modulates the AT model’s phase behavior on the UJ lattice as shown in Fig.~\ref {fig:lattice}. 
We further note that two-dimensional inhomogeneous lattices include the dice lattice as the simplest representative, which provides useful context for understanding the more complex UJ lattice and the associated critical phenomena that may emerge under structural inhomogeneity~\cite{VALDES2007551,26}.

Numerous spin systems have been explored on UJ lattices to characterize their critical properties. Traditionally, research in this area has focused primarily on critical points or lines, while the critical phase  remains underexplored, its characteristics in UJ lattices still require clarification. For example, the XY model~\cite {XYmodel_UJ}, clock model~\cite {trg_UJ}, and quantum spin models~\cite {spin_wave_UJ, hsb_UJ} on UJ lattices exhibit a critical phase (an intermediate state between FM and PM phases), while the pure antiferromagnetic Potts model~\cite {q_potts_xiang, deng2011finite} and the Baxter-Wu model~\cite {dingUJ} on UJ lattices only show critical points, with no evidence of a critical phase.

Even in structurally simple square lattices, the emergence of a critical phase is not universal and requires specific conditions. For example, the $q$-state clock model \cite{tnclock8, tn-clock5, NN-clock6} exhibits a critical phase only when $q > 4$; the three-state ferromagnetic Potts model~\cite{potts-vl4} and antiferromagnetic Potts model~\cite{Potts3}  must include an additional vortex term to manifest a critical phase.
Without considering such vortex energy terms, a critical phase can still emerge if the single-layer geometry is modified to a layered structure, for example, in studies of the four-layer antiferromagnetic Potts model~\cite{potts7}.
Similar lattice-dependent conditions apply to other systems: the frustrated \(J_1-J_2\) Ising model on kagome lattices also hosts critical phases~\cite{Cheng_Chen,Cheng_Chen_1}, though its underlying mechanism may differ from those of square lattice systems.

In any case, to our knowledge, the AT model in the UJ lattices has not yet been studied, and there has been no prediction of whether there is a critical phase.
In this paper, we systematically explore the AT model of the UJ lattice.  In particular, compared to the phase diagrams in square and other lattices \cite{worm}, we reveal a novel critical phase unique to the UJ lattice geometry for the AT model.
The critical phase we have discovered exhibits many interesting characteristics.
Firstly, in the critical phase, the algebraic decay of the correlation function with the size of the system is indirectly confirmed by measuring the algebraic decay of magnetization of coupled spins with size~\cite{dingUJ,log}.
Secondly, a new definition of magnetic susceptibility is proposed, yielding evidence that the phase transition between the critical phase and other phases belongs to the BKT transition, which is easier than using  the traditional susceptibility or Binder ratio.
Finally, and equally importantly, due to the inhomogeneity of the lattice, it is also found that the critical exponents $\eta$ of different sublattices are distinct.


The outline of our paper is as follows.
Sec.~\ref{sec:model} presents the lattice, method, and the measured quantities. The results about the phase diagram and phase transitions  are shown in Sec.~\ref{sec:resa}.
Conclusion and discussion are given in Sec.~\ref{sec:con}.

\section{Lattice, Method and Quantities}
\label{sec:model}

Figure~\ref{fig:lattice} presents a schematic of the AT model on a \( 4 \times 4 \) UJ lattice under periodic boundary conditions. To describe the lattice behavior in detail, we categorize all lattice sites within a layer into three sublattices: 0-sub-lattice sites with a coordination number of 4, and 1- and 2-sub-lattice sites (two distinct subsets) with a coordination number of 8. 
In Ref.~\cite{XYmodel_UJ}, Deng {\it et al.} labeled them as $A_4$, $B_8$ and $C_8$, and such a classification helps to explicitly distinguish the sublattices.


We perform Monte Carlo simulations using the Metropolis algorithm~\cite{Metropolis1953}. 
The spins $\sigma$, $\tau$, and $s = \sigma\tau$ are updated sequentially, with the other two spins fixed during each update. 
Each trial flip is accepted with probability 
\begin{equation}
P_{\text{acc}} = \min\left(1, e^{-\Delta E/(k_B T)}\right),
\end{equation}
where $\Delta E$ denotes the energy change of the system. 

The quantities to be measured by the Metropolis method~\cite{Metropolis1953} are as follows:

\begin{itemize} 
\item Magnetization 
\begin{equation}
M_{\alpha}^{\sigma} =  |\sum_{ i=1}^{N_\alpha} \sigma_{i}|, \label{eq:mf}
\end{equation}
where the subscript $\alpha$ takes values 0, 1, 2,  $t$ representing the 0th sublattice, the 1st sublattice, the 2st sublattices, and the total lattices. $N_\alpha$ is the number of spins on the lattices.
The  magnetization per-site  $m_{\alpha}^{\sigma}$ is defined by $ M_{\alpha}^\sigma/N_\alpha$. The superscript $\sigma$ represents $\sigma$-spins, which can be replaced by $\tau$ and $s$ to get $m_{\alpha}^{\tau}$, and $m_{\alpha}^s$ for the $\tau$-type spins and the  $s$-spins similarly.

\item  Binder ratio
\begin{equation}
Q = \frac{\langle M^{2} \rangle^{2}}{\langle M^{4}\rangle},\label{eq:br}
\end{equation}
where $M$ can be magnetization on different sub-lattices and different types of spins  such that  $Q_{\alpha}^s$ and $Q_{\alpha}^{\sigma}$ can be obtained.

\item Magnetic susceptibility
\begin{equation}
\chi = \frac{1}{NT} \left( \langle M^2 \rangle - \langle |M| \rangle^2 \right),
\label{eq:chi}
\end{equation}
where $M$ can be   $M_{\alpha}^{\sigma}$ and $M_{\alpha}^{s}$ to get $\chi_{\alpha}^{\sigma}$ and $\chi_{\alpha}^{s}$, $N$ can be $N_\alpha$ and   the temperature $T$ is set to be 1 in the simulations. Different values of $\alpha$ are assigned to different sublattices, 
enabling us to clearly separate and compare their magnetic susceptibility responses.

\item The correlation length \(\xi\) between spins is another marker to identify critical states. It is defined by the formula~\cite{potts7,17_1}:
\begin{equation}
    \xi = \frac{1}{2 \sin(k_m/2)} \sqrt{\frac{\langle m(\vec{0})^2 \rangle}{\langle m(\vec{k}_m)^2 \rangle} - 1}
    \label{eq:correlation_length}
\end{equation}
where \(k_m = 2\pi/L\) (with \(L\) denoting the system size) and \(\langle m(\vec{k})^2 \rangle\) is the $k$-dependent magnetization, defined as:
\begin{equation}
    \langle m(\vec{k})^2 \rangle = \left\langle \left| \frac{1}{N} \sum_{i=1}^N s_i \exp(i\vec{k} \cdot \vec{r}_i) \right|^2 \right\rangle
    \label{eq:magnetization}
\end{equation}

For a second-order phase transition or a BKT transition, \(\xi/L\) values for different system sizes \(L\) approximately intersect at the critical temperature~\cite{20_1,17_1,22_1,21_1}.
In the real simulation, the correlation length ratio of \(\xi^{\sigma}_t/L\) and  \(\xi^{s}_t/L\) are measured.
\end{itemize}

\section{Numerical Results}
\label{sec:resa}

Given the richness of the AT model on UJ lattices, this paper focuses solely on the phase diagram and critical phases for the case \( J > 0, K < 0\). We also provide the general phase diagram for \(K > 0\) for completeness.

\subsection{Phase Diagram and Typical Phases}
\label{sec:typicalphase}
\subsubsection{Phase Diagram}
\begin{figure}[htbp]
    \centering    \includegraphics[width=0.5\textwidth]{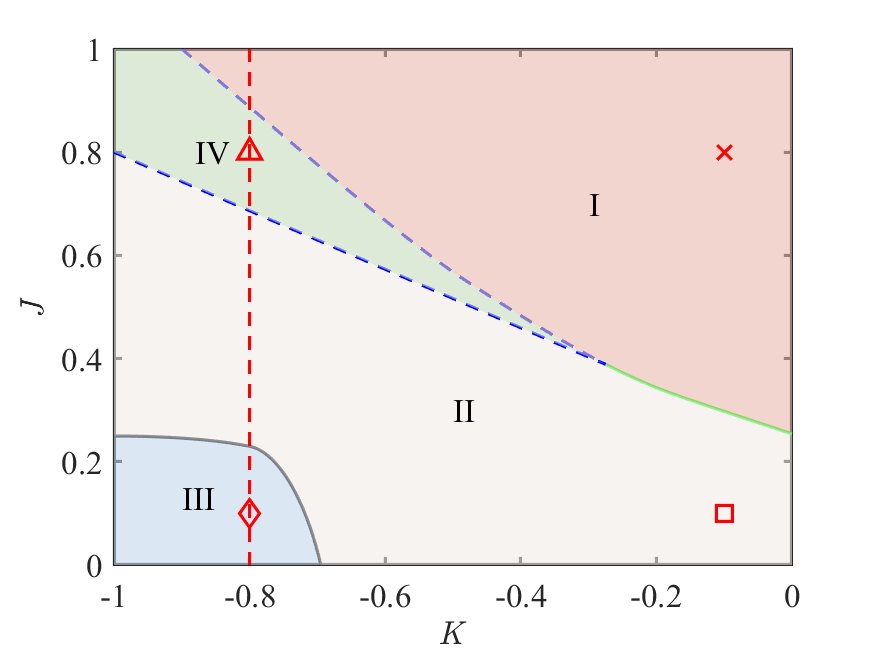}
    \caption{ Phase diagram of the AT model on the UJ lattice at fixed temperature \(T=1\). 
The diagram contains four distinct phases (I–IV): the  ferromagnetic phase, the  paramagnetic (PM) phase, the partially antiferromagnetic phase, and the critical phase, respectively. 
The red dotted line \(K=-0.8\) is included to illustrate the details of phase behavior and transitions along this cut. 
The four symbols mark representative points in each phase, with each corresponding to a specific parameter \(\vec{m}\) selected from the respective phase region. 
The blue dashed line denotes the BKT transition, the black solid line denotes the Ising transition, and the green solid lines mark other continuous phase transitions.}
    \label{fig:phase_+J}
\end{figure}

In Fig.~\ref{fig:phase_+J}, we present the phase diagram of the AT model in the UJ lattice in $K-J$ plane in the range $-1<K<0$ and $0<J<1$;
given the richness of the full phase diagram, only this parameter regime is focused on in the present work.
The lines in the diagram are schematic boundaries for visual guidance. 
The red dashed line is selected as the pathway along which the data are presented comprehensively, for a detailed illustration of the phases and transitions.
 
 The I and II  phases are consistent with their corresponding phases in the AT model on uniform lattices~\cite{worm}. Specifically, phase I is an FM phase in which all types of sub-lattices exhibit FM ordering. Phase II corresponds to a PM phase, with all spins on sub-lattices are in the PM phase.
 Phase III is characterized by a partial antiferromagnetic order in the $s$-spins. Among them, the spins on sublattices 1 and 2 exhibit an AFM phase, while the spins on sublattice 0 are in a PM phase. The $\sigma$-spins on all sublattices are  in the PM phase.

The IV phase is the emergent critical phase different from the previous works such as AT model on the uniform lattices~\cite{worm}. 
There exist numerous criteria and judgment standards for identifying its critical phase. For instance, the magnetization of the finite-size system decays algebraically with the system size following a power-law, which adheres to the formula: 
\begin{equation}
\left \langle m(L)\right \rangle \propto L^{-k},
\end{equation}
where $\left \langle m(L)\right \rangle$ denotes the magnetization of the system with size $L$, and $k$ represents the critical exponent characterizing the power-law decay behavior. Furthermore, both its Binder ratio  $Q(L)$ and correlation length ratio $\xi/L$ converge to finite values when $L\rightarrow \infty$.
There are also signs of the BKT phase transition in the transition types between the critical phase and other phase, such as the PM phase and the FM phase.

\subsubsection{Symmetry of sublattices}

The sublattice inhomogeneity of the lattice causes the order parameter to show significant differences from that of the uniform lattice. We define a vector $\vec{m}$ in the $x-y$ plane, and the distribution of this order parameter precisely reflects this inhomogeneity. The expression of $\vec{m}$ is given as follows:

\begin{equation}
\begin{tikzpicture}[baseline, scale=0.8] 
    \node[anchor=east] (vectors) {
        \begin{tikzpicture}[scale=0.8] 
            \draw[->, thick, red] (0,0) -- (0,1) node[left] {$\vec{e}_0$};
            \draw[->, thick, blue] (0,0) -- ({sqrt(3)/2}, -0.5) node[right] {$\vec{e}_1$};
            \draw[->, thick, green!60!black] (0,0) -- ({-sqrt(3)/2}, -0.5) node[left] {$\vec{e}_2$};
            \fill (0,0) circle (1pt) node[below] {$O$}; 
        \end{tikzpicture}
    };
    
    \node[anchor=west, xshift=1pt] (equation) at (vectors.east) {
        $\begin{split}
            \vec{m} &= m_0^s \, \vec{e}_0 + m_1^s \, \vec{e}_1 + m_2^s \, \vec{e}_2, \\
            \vec{e}_0 &= (0, 1), \\
            \vec{e}_1 &= \left( \sqrt{3}/2, -1/2 \right), \\
            \vec{e}_2 &= \left( -\sqrt{3}/2, -1/2 \right),
        \end{split}$
    };
\end{tikzpicture}
\end{equation}

\begin{figure}[t]
    \centering
\includegraphics[width=0.5\textwidth]{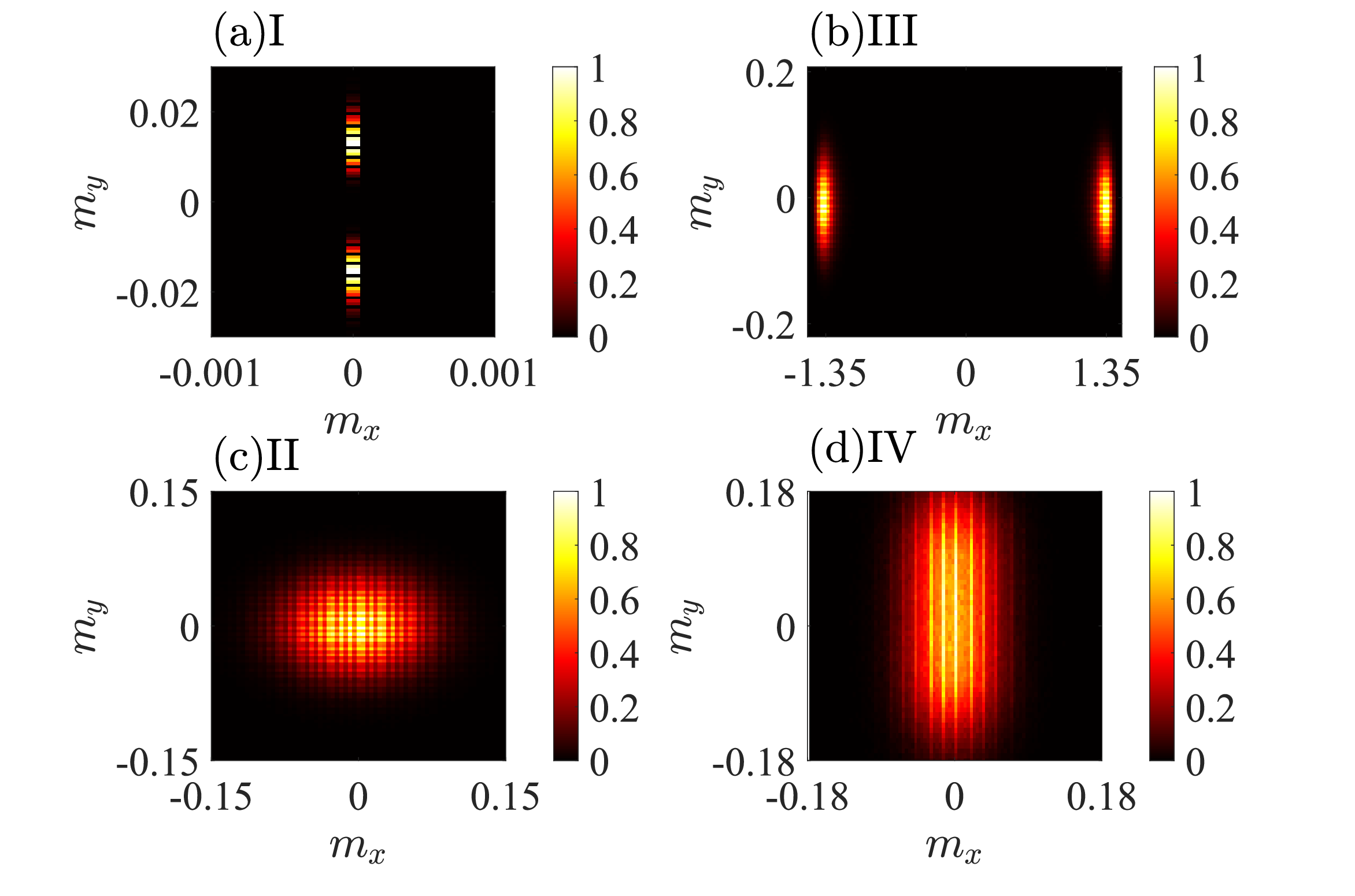}
    \caption{ Two-dimensional distribution  of $P(\vec{m})$ of $s$-spins in the four phases for $L=64$. The parameters are (a) phase I: $K=-0.1$,  $J=0.8$ (b) phase III: $J=0.1, K=-0.8$ (c) phase II: $J = 0.1$, $K=-0.1$ for the II phase (d) phase IV: $J = 0.8, K = -0.8$. The positions of the parameter points in the phase diagram are marked respectively with red  scrosses, diamonds, squares and triangles.}
\label{fig:O2}
\end{figure}
\noindent where \(m_0^s\), \(m_1^s\), and \(m_2^s\) are uniform magnetization in Eq.~\ref{eq:mf}  summing the spins of sublattice 0, sublattice 1, and sublattice 2, respectively.
The statistical histograms of $\vec{m}$ in different phases in the two-dimensional plane are plotted.  In Fig.~\ref{fig:O2} (a), ideally, i.e., $J\longrightarrow \infty$, in  phase I,  the  values of $m_0^s$, $m_1^s$, and $m_2^s$ should be 1. The non-zero distribution  $P(\vec{m})$ is   concentrated at the position of $\vec{m}=0$, and zero elsewhere $\vec{m}\ne 0$. In real finite $J$, i.e., $J=0.8$ and $K=-0.1$, two very narrow bright spots are observed,   symmetric about the horizontal axis. Specifically, the spot with $m_y<0$ satisfies $m_0^s = 1$ and $m_1^s = m_2^s = 0.97$, while the spot with $m_y>0$ corresponds to $m_0^s = -1$ and $m_1^s = m_2^s = -0.97$.
In Fig.~\ref{fig:O2} (b), for the phase III, the non-zero region of \(P(\vec{m})\) occurs where \(m_x \approx \pm 1.35\) at $J=0.1$ and $K=-0.8$, a value close to the ideal value $\pm \sqrt{3}$. This is because, in phase III, \(m_0^s \approx 0\) and \(m_1^s = -m_2^s \approx \pm 1\).
The sublattice 0 ($A_4$) has fewer neighboring sites compared to the sublattices 1 ($B_8$) and 2 ($C_8$), equivalent to having weak interactions, thus being in a PM state, therefore $m_0^s\approx 0$. Meanwhile, 
the $s$-spin on sublattices 1 and 2 exhibits an AFM  phase.
In Fig.~\ref{fig:O2} (c), for the phase II, $P(\vec{m})$  is a bright spot with elliptical shape rather than circular shape.
The shape is not rotational invariant, which means that the sublattice inhomogeneity is visible.
In Fig.~\ref{fig:O2} (d), for the phase IV, the distribution $P(\vec{m})$ is similar to that of the PM phase, and both do not have rotational invariance. 
The distribution in the directions of $m_x$ and $m_y$ is not completely symmetric, which can reflect the inhomogeneity of the sublattices $A_4$ and $B_8$ or $C_8$.

\subsubsection{The snapshots}

\begin{figure}[htbp]
\centering
\includegraphics[width=0.48\textwidth]{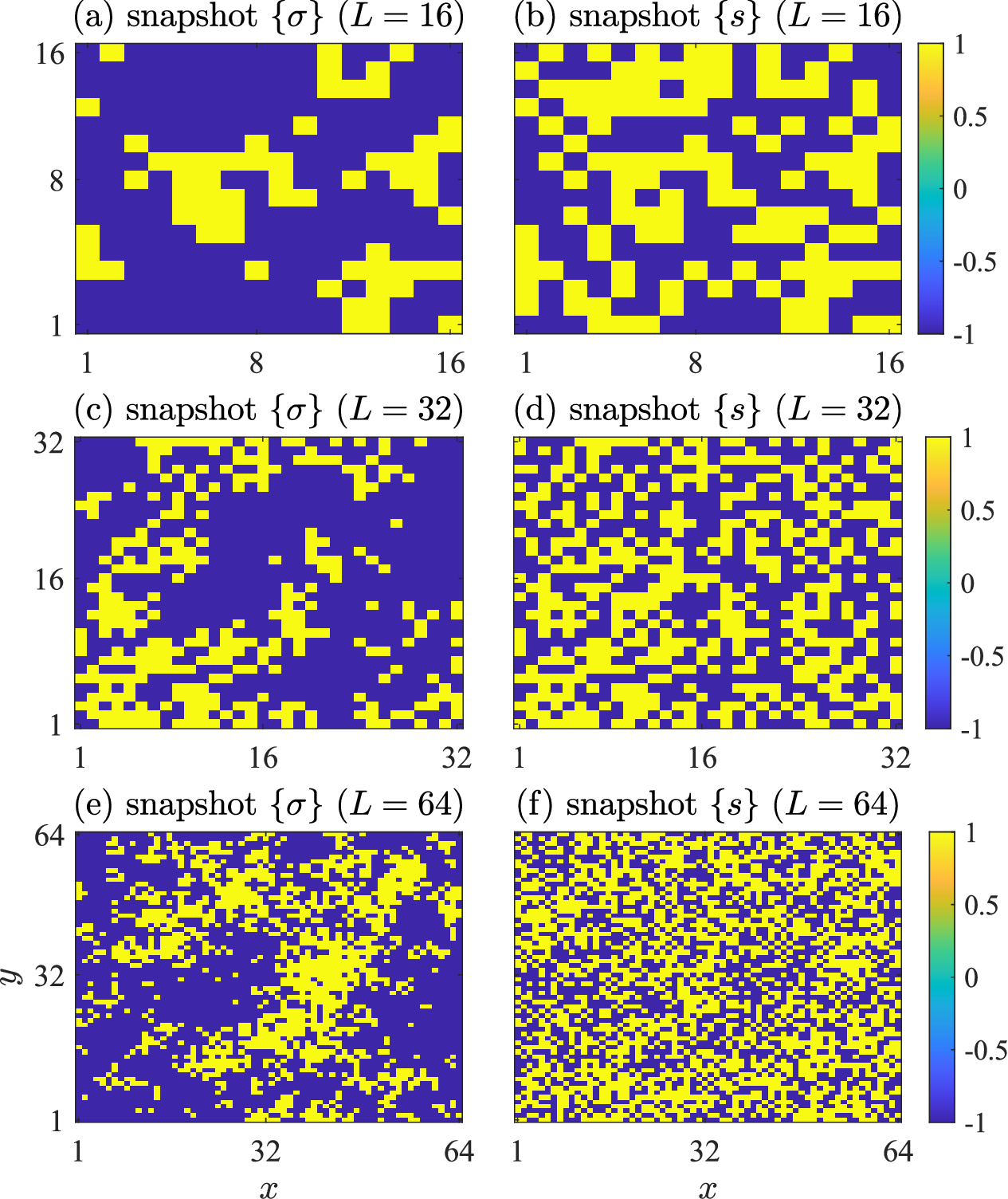}
\caption{The snapshots of $\{\sigma\}$-sipns and $\{s\}$-sipns at $L=16，32，64$ in the critical phase IV: (a) $\{\sigma\}$-spins; (b) $\{s\}$-spins at $J = 0.8$ and $K = -0.8$, respectively.}
\label{fig:JGYZ}
\end{figure}


In Fig.~\ref{fig:JGYZ}, the snapshots presented here illustrate the spatial spin configurations characteristic of the critical phase. Spin $+1$ and spin $-1$ are represented by yellow and blue, respectively.
Figure~\ref{fig:JGYZ}(a) displays a typical configuration of the $\{\sigma_{i}\}$ spins at $J=0.8$, $K=-0.8$ for lattice size $L=16$.
Although the instantaneous magnetization $m^{\sigma}$ appears large and non‑zero, with a clear dominance of spin $-1$, such a single finite‑size snapshot is not sufficient to identify a bulk ferromagnetic phase.
Instead, the systematic size dependence of the averaged magnetization $\langle m^{\sigma}\rangle$ shows a power‑law scaling with $L$, which signals that the $\sigma$ spins reside in a quasi‑long‑range ordered critical phase, not a true ferromagnetic phase.

In contrast, the $s$-spin configuration in Fig.~\ref{fig:JGYZ}(b) exhibits a weak instantaneous magnetization $m^s$, with a seemingly random distribution of yellow and blue spins that could be misinterpreted as a paramagnetic phase.
However, the \emph{averaged} magnetization $\langle m^s\rangle$ also follows a power‑law dependence on system size $L$, confirming that the $s$ spins are likewise in the quasi‑long‑range ordered critical phase.

These observations convey a key message: instantaneous snapshots and single‑configuration magnetizations can be misleading in finite systems.
To make this point rigorous, we provide in Fig.~\ref{fig:JGYZ} additional configurations at $J = 0.8$, $K = -0.8$ for system sizes $L = 16, 32, 64$.
We select representative configurations whose magnetizations are closest to the ensemble average, with the corresponding values given below:
\begin{equation}
\begin{aligned}
L&=16: \ m_\sigma=0.5156,\ m_s=0.0781, \\
L&=32: \ m_\sigma=0.4375,\ m_s=0.0723, \\
L&=64: \ m_\sigma=0.3760,\ m_s=0.0356.
\end{aligned}
\end{equation}
Together with the size scaling of the averaged magnetization, these data confirm that the phase is critical quasi‑long‑range ordered, not ferromagnetic or paramagnetic.

\subsection{The detail of the phases and phase transitions}

\begin{figure}[t]
\centering
\includegraphics[width=0.5\textwidth]{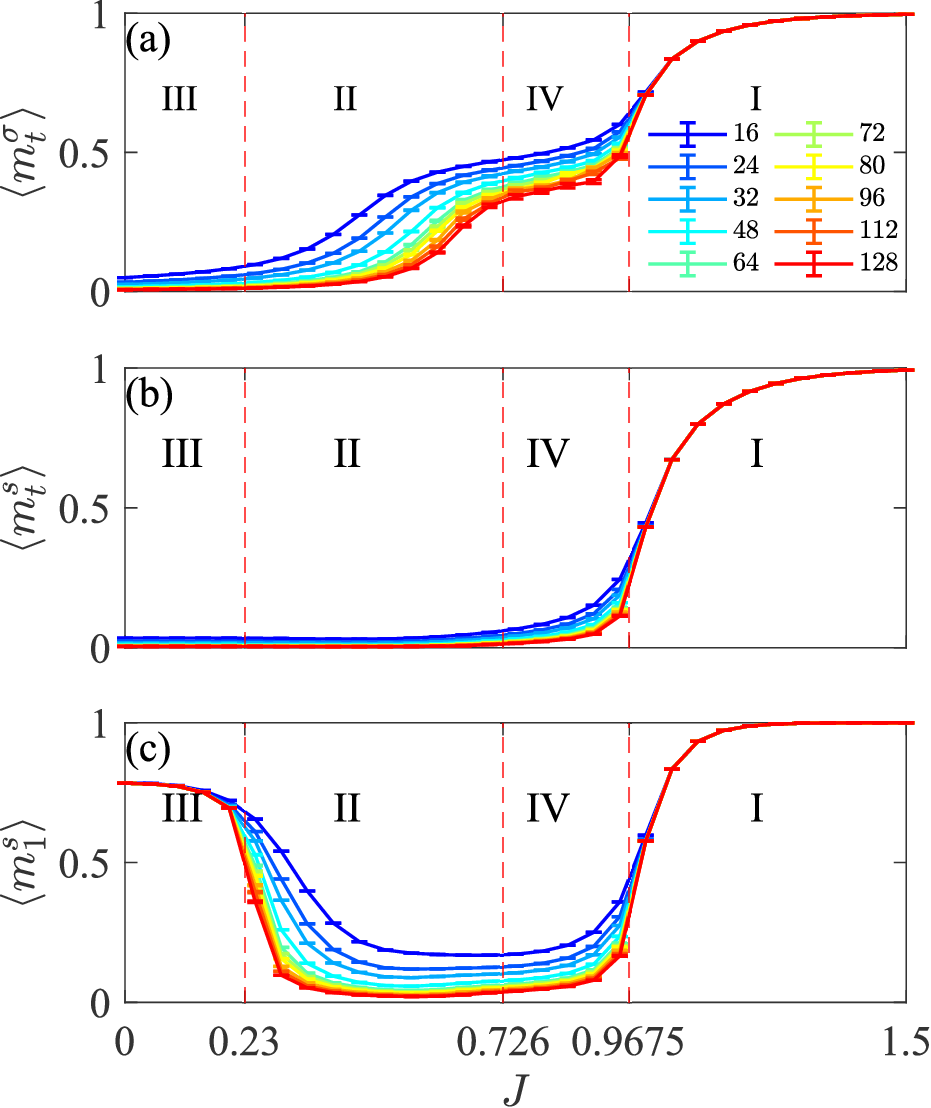}
    \caption{The details of the phase transition from phase I to phase IV are along the red dashed line in Fig.~\ref{fig:phase_+J}, where $K=−0.8$. (a) $\left \langle m_t^{\sigma}\right \rangle$ (b) $\left \langle m_t^s\right \rangle$ (c) $\left \langle m_{1}^s\right \rangle$. The red dashed line and the thick green line represent the phase transition boundaries.}
\label{fig:M}
\end{figure}

\subsubsection{magnetization and magnetic susceptibility}
In Figs.~\ref{fig:M}, the magnetism curves of $\langle m^{\sigma}_t\rangle$, $\langle m^s_t\rangle$, and $\langle m^s_1\rangle$ versus $J$ are shown along the dashed line in Fig.~\ref{fig:phase_+J} for lattice sizes $L=16$ to $128$.

In the range $0<J<0.23$, the system is in phase III, where $\langle m^{\sigma}_t\rangle=0$, $\langle m^s_t\rangle=0$, and $\langle m^s_1\rangle\neq0$. Phase III is a partial AFM phase, characterized by $\langle m^{s}_0\rangle=0$ (indicating the spins in the 0-th sublattice are in the PM phase) and $\langle m^{s}_1\rangle\neq0$, $\langle m^{s}_2\rangle\neq0$ with mutually opposite spin directions.
Two fundamental reasons account for the emergence of this partial AFM phase: 
Firstly, when $K<0$, the system exhibits AFM interaction characteristics, and since $|K|>|J|$, the $K$-driven interaction dominates over the $J$-driven interaction. 
Secondly, in the UJ-lattice, the 0-th sublattice has only 4 neighboring sites (low coordination number), resulting in weak average spin-spin interactions; in contrast, sublattices 1 and 2 each have 8 neighboring sites (higher coordination number), which enhances their effective spin-spin interactions. Thus, AFM order emerges in sublattices 1 and 2, while the 0-th sublattice remains in the PM phase. 

The existence of this phase directly manifest the entropic selection effect in this system~\cite{26,PhysRevB.66.024422}: the high-coordination sublattice, leveraging its entropic advantage, develops antiferromagnetic order under thermal fluctuations, whereas the low-coordination sublattice persists in a disordered state. 
The observed entropic selection arises directly from the free energy competition \(F = U - TS\). At finite temperature, the system favors a partially ordered state with higher entropy over a fully ordered state with lower energy, because the entropic term \(-TS\) dominates the free energy and stabilizes the high‑entropy ordered phase.

In the range $0.23<J<0.717$, the system is in the PM phase, as $\langle m^{\sigma}_t\rangle$, $\langle m^s_t\rangle$, and $\langle m^s_1\rangle$ are all zero. 
For $J>0.975$, $\langle m^{\sigma}_t\rangle$, $\langle m^s_t\rangle$, and $\langle m^s_1\rangle$ are all non-zero, indicating the system is in FM phase. 
In the range $0.717<J<0.975$, the system is in the critical phase: unlike the PM phase  and FM phase, the magnetization in this range shows a power-law dependence on lattice size $L$ (detailed evidence is presented in subsequent sections).

\begin{figure}[htb]
    \centering
    \includegraphics[width=0.5\textwidth]{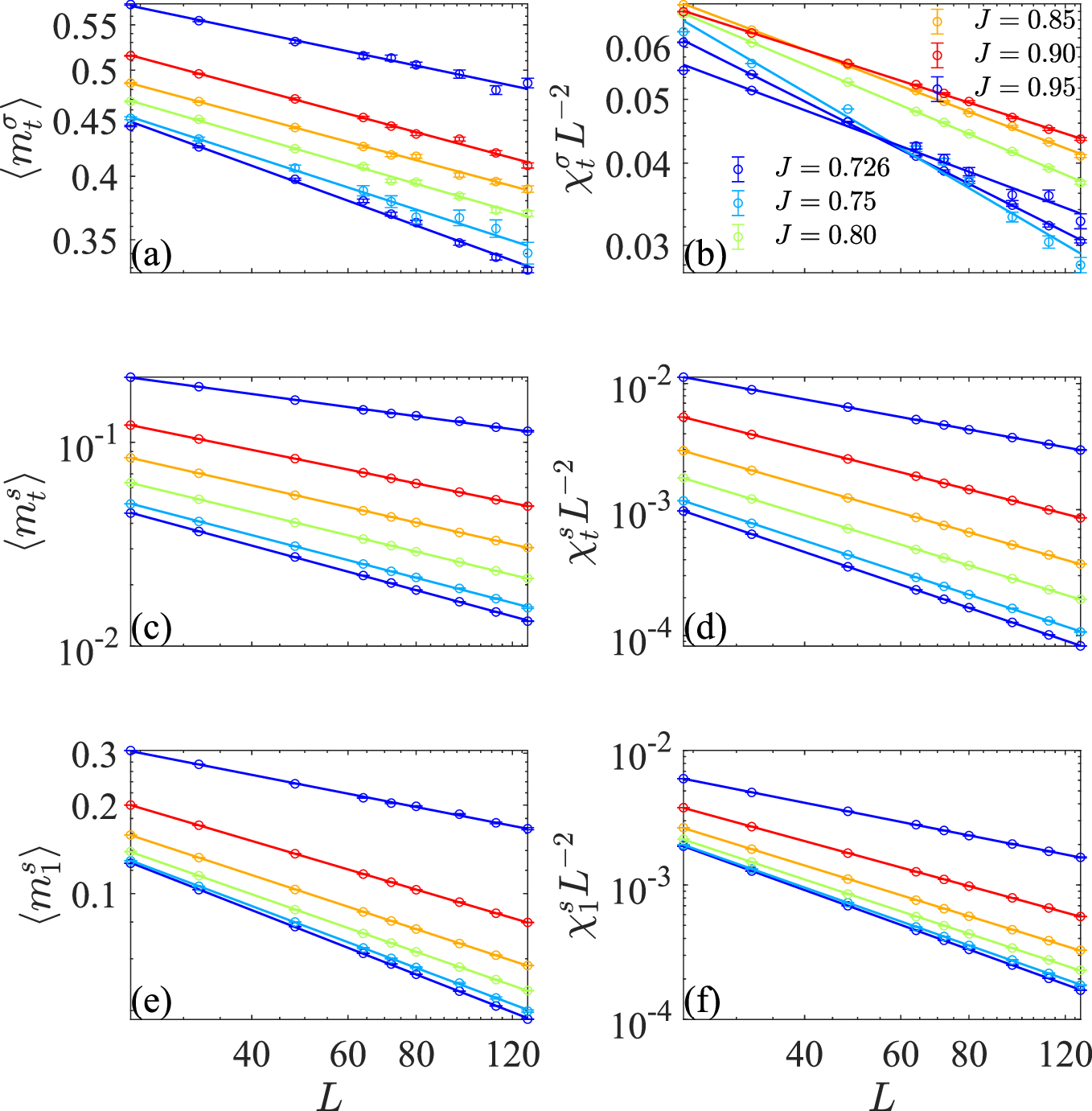}
    \caption{The left columns (a), (c), and (e) are loglog plots of $\left \langle m\right \rangle$ versus $L$, while the right columns (b), (d), and (f) are loglog plots of $\chi L^{-2}$ versus $L$. The parameters are $J=0.75-0.9$ with $K=-0.8$.}
\label{fig:dieluo_M_X}
\end{figure}

In Fig.~\ref{fig:dieluo_M_X}, the data in the left column correspond to log-log plots of $\left \langle m\right \rangle$ versus $L$, whereas those in the right column represent log-log plots of $\chi 
 L^{-2}$ versus $L$, with the system size $L$ ranging from 24 to 112.

As described in Refs.~\cite{Cheng_Chen,Cheng_Chen_1}, the magnetization and magnetic susceptibility at the critical point satisfy the following scaling relations:
\begin{equation}
\begin{split}
\langle m\rangle &= L^{-\eta/2} \, \mathcal{F}_{\langle m\rangle} \left( \xi / L\right), \\
\chi &= L^{2-\eta} \,\mathcal{F}_\chi\left( \xi / L \right),
\end{split}
\label{eq:M_X_log}
\end{equation}
where $\eta$ is the magnetic anomalous dimension, and $\mathcal{F}_{\langle m\rangle}$ and $\mathcal{F}_\chi$ are universal analytic scaling functions. As commonly adopted in literature~\cite{ChenTao.111.094201, 7Parisen_Toldin_2009}:, these scaling functions can be expressed as:
\begin{equation}
\begin{split}
\langle m\rangle &= L^{-\eta/2}(a + bL^{-\omega}), \\
\chi &= L^{2-\eta}(a + bL^{-\omega}),
\end{split}
\label{eq:scaling_fit}
\end{equation}
where $\omega$ is the correction-to-scaling exponent (set to 1 in this work), and $a$, $b$ are non-universal fitting parameters. 
In the fitting process, fixing $b=0$ leads to significantly inconsistent values of $\eta$ when fitting $\langle m\rangle$ and $\chi$ separately; by contrast, including $b$ in the fitting yields consistent $\eta$ values, with deviations within a factor of three in the error bars. 
Furthermore, to avoid underestimating the fitting uncertainty, we incorporate the inherent error of each data point into the fitting procedure.
\begin{table*}[htbp]
  \centering
  \caption{This table presents the critical exponent $\eta$ for different values of $J$ at $K=-0.8$. The superscripts $\sigma$ and $s$ denote different spins. The exponents are obtained by fitting the magnetization $\left \langle m\right \rangle$ and the magnetic susceptibility $\chi$. The numbers in parentheses represent the statistical errors (3 times the standard deviation).}
  \setlength{\tabcolsep}{9pt}
  \begin{tabular}{cccccccl}  
    \hline \hline
     & $J$ & 0.726 & 0.75 & 0.8 & 0.85 & 0.9 & 0.95\\ 
    \hline 
    $\left \langle m_{t}^{\sigma}\right \rangle$ & $\eta$ & 0.50(2) & 0.34(9) & 0.38(8) & 0.27(6) & 0.27(2) & 0.21(1) \\

    $\chi_{t}^{\sigma}$ & $\eta$ & 0.49(1) & 0.47(7) & 0.35(3) & 0.32(3) & 0.28(1) & 0.35(6)\\
    
    $\left \langle m_{0}^s\right \rangle$ & $\eta$ & 1.88(3) & 1.71(3) & 1.85(3) & 1.77(2) & 1.12(5) & 1.0(6)\\
    
    $\chi_{0}^s$ & $\eta$ & 1.85(3) & 1.72(2) & 1.85(3) & 1.77(3) & 1.20(3) & 0.73(6)\\
    
    $\left \langle m_{1}^s\right \rangle$ & $\eta$ & 1.65(6) & 1.39(2) & 1.28(3) & 1.18(6) & 1.10(6) & 1.0(6) \\
    
    $\chi_{1}^s$ & $\eta$ & 1.53(3) & 1.39(2) & 1.30(6) & 1.21(3) & 1.10(1) & 0.78(2)\\

    $\left \langle m_{t}^s\right \rangle$ & $\eta$ & 1.57(3) & 1.38(3) & 1.27(3) & 1.18(6) & 1.10(6) & 1.0(6) \\
    
    $\chi_{t}^s$ & $\eta$ & 1.47(6) & 1.38(3) & 1.29(6) & 1.21(1) & 1.10(1) & 0.78(2)\\
    
    \hline \hline
  \end{tabular}
  \label{tab:my_table}
\end{table*}

Table~\ref{tab:my_table} lists the values of $\eta$ obtained by fitting $\langle m\rangle$ and $\chi$. These two physical quantities are from the $s$-spins and $\sigma$-spins of the total lattice and sublattices, respectively. This observation is also intended to investigate whether the inhomogeneity of the sublattices can induce different power-law decay exponents.

For the $\sigma$-spins, only the fitted $\eta$ values of the total lattices are presented in the table. This is because these values show no statistically significant differences within the error bars from the spins on the sublattices. Perhaps higher-precision calculations could reveal differences in $\eta$ values between different sublattices for $\sigma$-spins. The $\eta$ values obtained in our calculations range from 0.21 to 0.50, which are significantly lower than the fitted $\eta$ for the $s$-spins. This can be attributed to the following physical mechanisms: the $\sigma$-spins interact via ferromagnetic $J$ couplings, leading to a tendency for parallel alignment among spins; consequently, the magnetization $\langle m^{\sigma}\rangle$ decays slowly with increasing system size. In contrast, the $s$-spins are subject to antiferromagnetic interactions and reside on a frustrated lattice, which gives rise to high degeneracy and large entropy. As a result, the magnetization of $s$-spins decays rapidly with the lattice sizes.

For the $s$-spins, the fitted $\eta$ values exhibit obvious differences among different sublattices. Specifically, the fitted $\eta$ values of sublattice 0 are larger than those of sublattice 1 and the total lattice. This phenomenon arises because the spins in sublattice 0 have only 4 neighbors, whereas those in sublattice 1 have 8 neighbors: weaker inter-spin interactions in sublattice 0 reduce the stability of long-range spin correlations, leading to a faster decay of magnetization. Meanwhile, the fitted $\eta$ value of sublattice 1 shows no noticeable difference from that of the total lattice within fitting errors, implying that higher-precision simulations would be required to identify any subtle distinctions.

\begin{figure}[t]
    \centering
\includegraphics[width=0.5\textwidth]{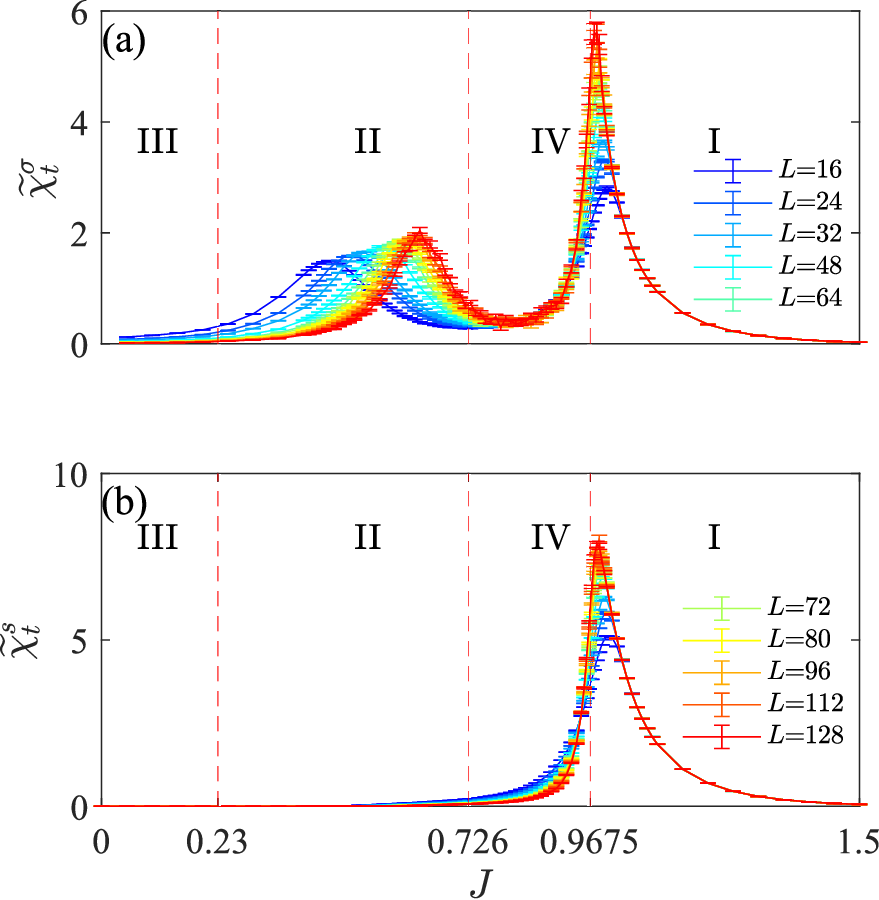}
\includegraphics[width=0.5\textwidth]{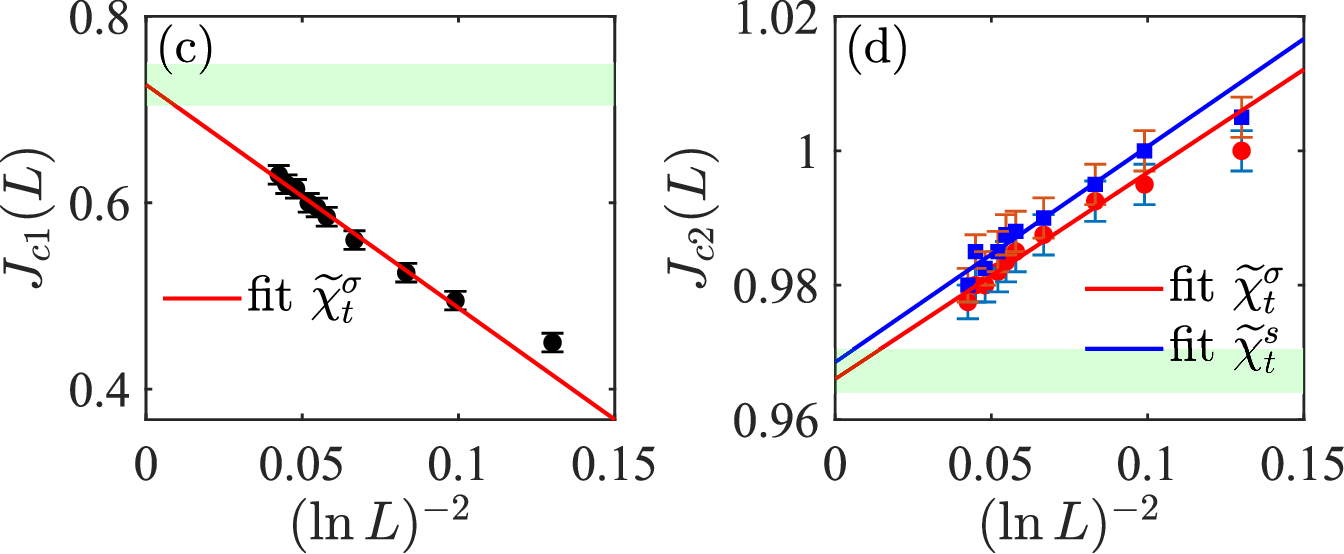}
    \caption{Details and evidence for the BTK phase transitions. (a) $\widetilde{\chi}_t^{\sigma}$ versus $J$ (b) $\widetilde{\chi}_t^{s}$ versus $J$ (c) $J_{c1}(L)$ versus $(\ln L)^{-2}$ by fitting $\widetilde{\chi}_t^{\sigma}$ (d)  $J_{c2}(L)$ versus $(\ln L)^{-2}$ by fitting $\widetilde{\chi}_t^{\sigma}$ and $\widetilde{\chi}_t^{s}$.}
    \label{fig:chipeaks}
\end{figure}

\subsubsection{new defined magnetic susceptibility and BKT transition}
Despite the identification of the critical phase, the nature of the phase transitions from this phase to PM phase and to the FM phase remains unconfirmed. The quasi-susceptibility \(\tilde{\chi} = d\langle m \rangle / d\lambda\) is a response function of the magnetization with respect to the control parameter \(\lambda\) (e.g., the coupling constants \(J\) or \(K\)), generalizing the standard susceptibility \(\chi = d\langle m \rangle / dH\), and follows the same spirit as the non-analyticity-based response functions widely used to detect quantum phase transitions, where derivatives of the ground-state energy with respect to a tuning parameter are employed to identify the order of the transition~\cite{sachdev1999quantum}. To verify whether these transitions belong to the BKT type, we leverage the variation of $\langle m_t^\sigma\rangle$ and $\langle m_t^s\rangle$ shown in Figs.~\ref{fig:M} to define two quasi-susceptibilities (analogous to the standard susceptibility $\chi = \partial\langle m\rangle/\partial h$, but differentiated with respect to coupling parameters $J/K$ for phase transition analysis):
\begin{equation}
\begin{split}
\widetilde{\chi}_t^\sigma &= \frac{\text{d}\langle m_t^\sigma\rangle}{\text{d}J} = \frac{\langle H_J\rangle\langle m_t^\sigma\rangle - \langle H_J m_t^\sigma\rangle}{J},\\
\widetilde{\chi}_t^s &= \frac{\text{d}\langle m_t^s\rangle}{\text{d}K} = \frac{\langle H_K\rangle\langle m_t^s\rangle - \langle H_K m_t^s\rangle}{K},
\end{split}
\label{eq:dmdj}
\end{equation}
where the Hamiltonians $H_J$ and $H_K$ are given by:
\begin{equation}
\begin{split}
H_J &= -J\sum_{\langle i,j\rangle }(\sigma_i \sigma_j + \tau_i \tau_j),\\
H_K &= -K\sum_{\langle i,j\rangle } s_i s_j.
\end{split}
\end{equation}
These expressions for the coupling susceptibilities, Eq.~\ref{eq:dmdj}, are derived directly from the standard canonical ensemble formalism presented in statistical mechanics textbooks.

Figure \ref{fig:chipeaks}(a) plots $\widetilde{\chi}_t^\sigma$ versus $J$ for lattice sizes $L=16, 32, 64, 128$. 
At the transition point between the critical phase and the PM phase (hereafter denoted as $J_{c1}(L)$), the finite-size peaks of $\widetilde{\chi}_t^\sigma$ shift slowly to the right as $L$ increases, a hallmark of BKT transitions, which obey the scaling relation~\cite{Weigel_2005_F_model,Cheng_Chen,Cheng_Chen_1}:
\begin{equation}
J_{c1}(L) = J_{c1}(\infty) + \frac{b_1}{(\ln L)^2},
\end{equation}
where $J_{c1}(\infty)$ is the critical coupling in the thermodynamic limit.
This is derived from the solution of the linearized renormalization group  equations for the two-dimensional  XY model near the BKT transition~\cite{Bramwell_1993,PhysRevB.49.8811}. Fitting the peak position data to this relation yields  $J_{c1}(\infty) = 0.726(5)$.

\begin{table}[htbp]
  \centering
  \caption{Finite-size scaling results for the critical coupling and slope under varying minimum system sizes $L_{\text{min}}$, with corresponding fitting quality quantified by $\chi^2/DF$.}
  \setlength{\tabcolsep}{9pt}
  \begin{tabular}{cccc}  
    \hline \hline
    
     $L_{min}$ & $\chi^2/DF$ & $J_{c1}$ & $b_1$\\ 
    \hline 
    $16$ & 0.717962 & 0.7096(7) & -2.0(1) \\
    $24$ & 0.191116 & 0.726(5) & -2.40(8)\\
    $32$ & 0.0919517 & 0.737(5) & -2.58(8)\\
    $48$ & 0.0231693 & 0.750(4) & -2.85(7)\\
    \hline \hline
  \end{tabular}
  \label{tab:jc1}
\end{table}
In table \ref{tab:jc1}, we list the fitting outcomes corresponding to different minimum system sizes \(L_{\text{min}}\), including the reduced chi-squared per degree of freedom \(\chi^2/DF\), the estimated critical coupling \(J_{c1}\) with statistical errors, and the critical slope with error bars. As the minimum system size \(L_{\text{min}}\) increases from 16 to 48, with fixed \(L_{\text{max}} = 128\), the value of \(\chi^2/DF\) decreases monotonically, indicating improved quality and reliability of the scaling fits.

Although the reduced chi-squared \(\chi^2/DF\) appears closest to unity for the fit including \(L_{\text{min}} = 16\), a visual inspection in Fig. \ref{fig:chipeaks}(c) reveals that the data point at \(L = 16\) clearly deviates from the asymptotic linear scaling behavior expected in the thermodynamic limit. This indicates that the smallest system size is still dominated by strong finite-size corrections and has not yet entered the scaling regime. Consequently, we discard \(L = 16\) from our analysis. Among the remaining fits, the case with \(L_{\text{min}} = 24\) yields the \(\chi^2/DF\) value closest to unity, providing the most reliable estimate of the critical coupling \(J_{c1} = 0.726(5)\), which we adopt in the following analysis.

For the transition between the critical phase and the FM phase (denoted as $J_{c2}(L)$), $\widetilde{\chi}_t^s$ curves are shown in Fig.~\ref{fig:chipeaks}(c). Although the peak positions of $\widetilde{\chi}_t^s$ show good convergence with increasing $L$, they still satisfy the BKT scaling relation:
\begin{equation}
J_{c2}(L) = J_{c2}(\infty) + \frac{b_2}{(\ln L)^2},
\label{eq:eqjcl}
\end{equation}
Fitting $\widetilde{\chi}_t^s$ and $\widetilde{\chi}_t^\sigma$ peak positions separately gives $J_{c2}(\infty) = 0.966(2)$ and $0.969(2)$, respectively, values consistent within their error bars, confirming the universality of the critical coupling for this transition.

\subsubsection{The behaviors of Binder ratio and correlation length}

\begin{figure}[t]
    \centering
\includegraphics[width=0.5\textwidth]{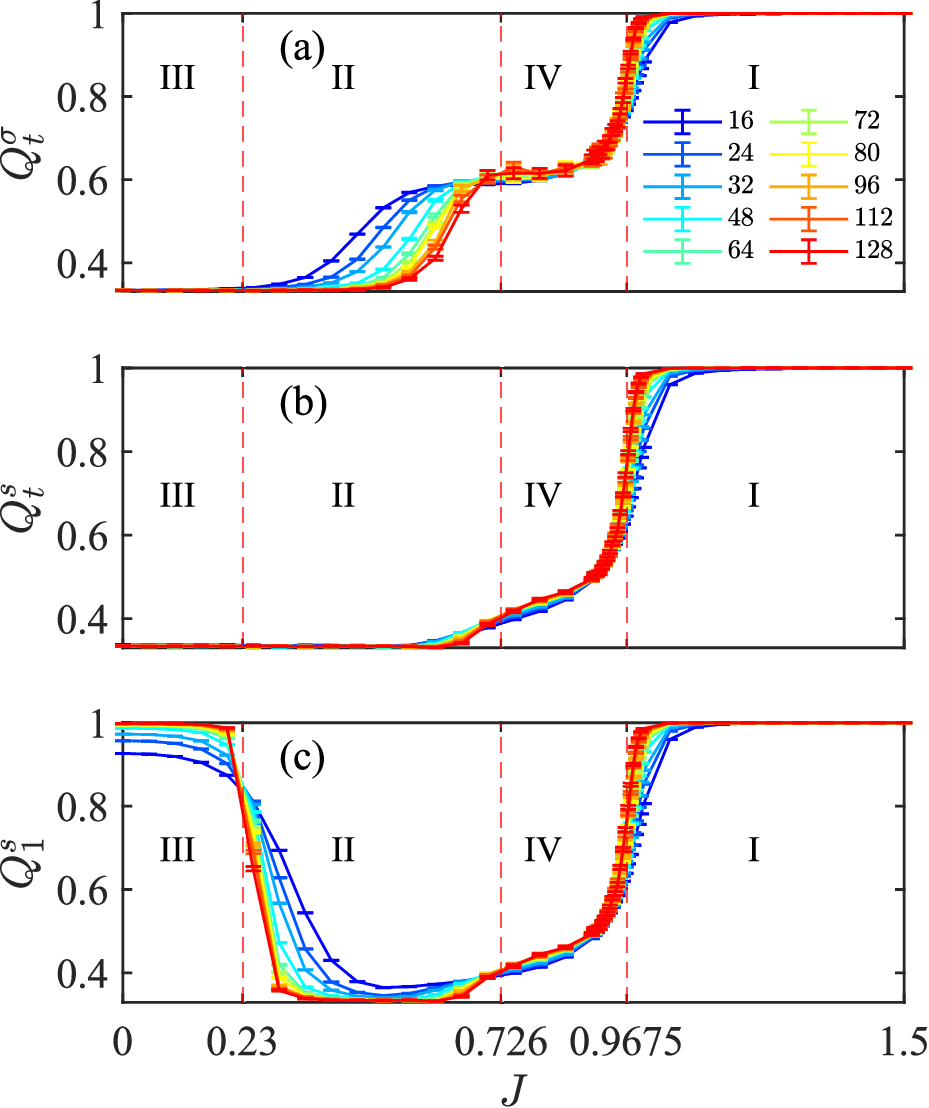}
    \caption{The details of the Binder ratio along the $K=-0.8$ (a) $Q^{\sigma}_t$ versus $J$ (b) $Q^{s}_t$ versus $J$ (c) $Q^{s}_1$ versus $J$.}
\label{fig:J=-K}
\end{figure}

\begin{figure}[t]
    \centering
\includegraphics[width=0.48\textwidth]{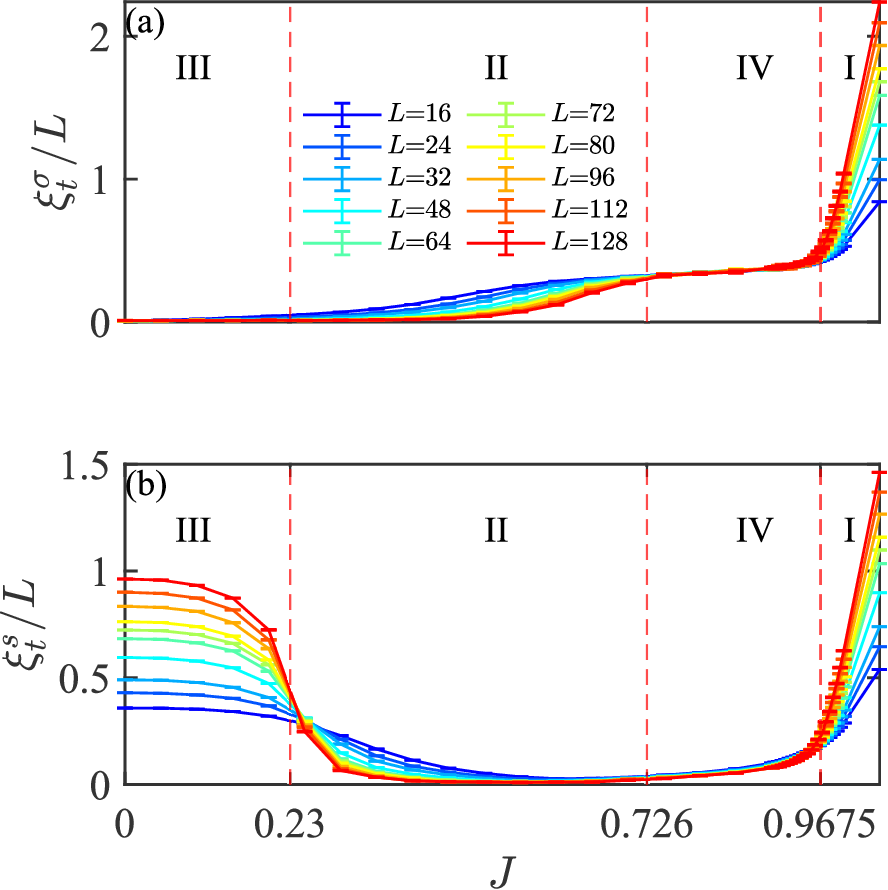}
    \caption{ The details of the correlation length along the $K=-0.8$ (a) $\xi_t^{\sigma}$ versus $J$  (b) $\xi_t^{s}$ versus $J$. The lattice  sizes range from \( L = 16 \) to \( L = 128 \).} 
    \label{fig:q}
\end{figure}
In Figs.~\ref{fig:J=-K} (a) and (b), the Binder ratio $Q_t^\sigma$ for $\sigma$-spins and $Q_t^s$ for $s$-spins are plotted as functions of $J$. In the range $0<J<0.23$, $Q_t^\sigma \approx 1/3$ and $Q_t^s \approx 1/3$. This behavior comes from the paramagnetic phase, where the total magnetization $M$ follows a Gaussian distribution. The Binder ratio is defined as $Q =  \langle M^2 \rangle^2/\langle M^4 \rangle$. Substituting the Gaussian integral results:
\begin{align}
\int_{-\infty}^{\infty} M^2 \frac{1}{\sqrt{2\pi}} e^{-\frac{M^2}{2}} dM &= 1, \\
\int_{-\infty}^{\infty} M^4 \frac{1}{\sqrt{2\pi}} e^{-\frac{M^2}{2}} dM &= 3,
\end{align}
we get $Q = 1/3$, which matches our numerical results.

In the range $0.23<J<0.717$, the system stays in the PM phase and the Binder ratios tend to converge to $1/3$ with increasing system size.
In the range $0.717<J<0.975$, the $Q_t^\sigma$ vs. $J$ curve shows a small plateau around $0.6$. For $L=16$, we fit the magnetization distribution $P(M)$ with two superimposed Gaussian functions and obtain a Binder ratio of ~0.5. This indicates the magnetization distribution in the critical phase is broadened, differing from the single-peak Gaussian distribution that gives $Q=1/3$. We also test a rectangular distribution for $P(M)$ and get  $Q_t^\sigma=5/9\approx0.56$, which is closer to the measured value $0.62(3)$.

In Fig.~\ref{fig:J=-K} (c), the boundary between phase II and phase III is extracted using the finite-size scaling formula:
\begin{equation}
Q = Q_0 + e_1(J - J_c)L^{y_t} + e_2(J - J_c)^2 L^{2y_t} + f_1 L^{-\omega},
\label{eq:q}
\end{equation}
where $J_c$ is the critical coupling, $y_t$ is the thermal critical exponent, and $\omega$ accounts for finite-size corrections. Fitting this formula to our data yields the critical point $J_c=0.23$.

This sensitivity to phase transitions also enables the Binder ratio to determine the BKT phase transition point, e.g., the boundary between phase IV and phase II. The proposed scheme is as follows:
For the $Q_{t}^{\sigma}$ curves (where finite-size effects are obvious), one can select a fixed value $Q_{t\sigma}^{*}$ (e.g., $0.5$) and plot its horizontal reference line. The values of $J$ corresponding to the intersections of this horizontal line and the $Q_{t}^{\sigma}$ curves are defined as the pseudocritical points $J_{c}(L)$. Finally, by fitting the $J_{c}(L)$ data for systems of different sizes with reference to Eq.~\ref{eq:eqjcl}, one can obtain the critical coupling strength in the thermodynamic limit, $J_{c}(\infty)$. Since we have already determined the BKT phase transition point using the quasi-susceptibility, this scheme is reserved for future investigation.

To further characterize the critical phases, we examine the behavior of the correlation length ratios $\xi_t^\sigma/L$ (for $\sigma$-spins) and $\xi_t^s/L$ (for $s$-spins), as shown in Fig.~\ref{fig:q}. 
In phase III, $\xi_t^\sigma/L = 0$. The $\sigma$-spin correlations are short-ranged and disappear in the thermodynamic limit because $J$ is no longer dominant. On the other hand, $\xi_t^s/L$ diverges. This matches the appearance of a partial AFM phase in $s$-spins, as $K$ takes the dominant role. In critical phase IV, $\xi_t^\sigma/L$ and $\xi_t^s/L$ converge to finite values in the thermodynamic limit. This signals a signature of critical phase.  
In the phase I, both ratios $\xi_t^\sigma/L$ and $\xi_t^s/L$ diverge, which is a direct consequence of true long-range spin order.
In phase II, both ratios vanish, indicating that the phase is the PM phase.

\begin{figure}[t]
    \centering
\includegraphics[width=1\linewidth]{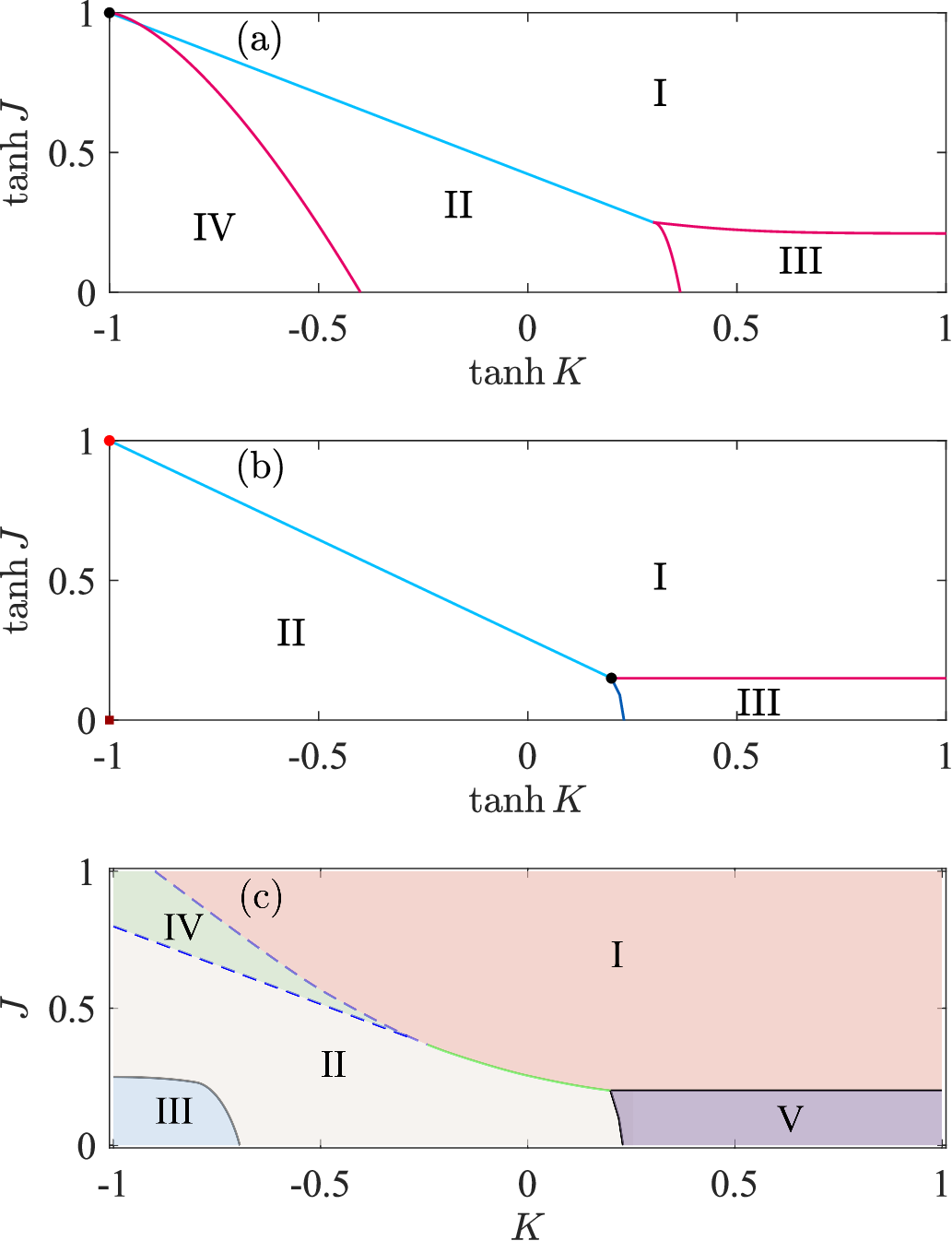}
  \caption{Phase diagrams of the  AT model for \(J>0\) on (a) the square lattice, (b) the triangular lattice~\cite{worm}, and (c) the  UJ lattice in present work.}
    \label{fig:AT_total}
\end{figure}
\subsection{Comparison of phase diagrams on square, triangular, and Union-Jack lattices}
Although we have focused our discussion on the phase diagram of the AT model on the UJ lattice for $J>0$ and $K<0$, we have not yet presented or reviewed the known phase diagrams of the AT model on conventional uniform lattices such as the square lattice, nor on frustrated lattices such as the triangular lattice\cite{worm}. Furthermore, the full phase diagram for $K>0$ has also not been included in the main text.

Figure \ref{fig:AT_total} (a) shows the phase diagram of the AT model on a square lattice with $J > 0$. The horizontal axis represents $\tanh K$ ranging from $-1$ to $1$, and the vertical axis represents $\tanh J$ ranging from $0$ to $1$. The system exhibits four phases, labeled I, II, III, and IV.
The behavior of $\langle\sigma\rangle$, $\langle\tau\rangle$, and $\langle\sigma\tau\rangle$ for each phase is detailed in Table~\ref{tab:phases_AT}.
Figure~\ref{fig:AT_total}(b) shows the phase diagram of the AT model on the triangular lattice for $J>0$, where only three stable phases exist.
In contrast to the union-jack lattice, neither the square nor the triangular lattice supports an emergent critical phase.

Figure~\ref{fig:AT_total}(c) displays the phase diagram of the AT model on the UJ lattice for $J>0$, where five distinct phases emerge, including the unique critical phase (phase IV).
In sharp contrast to the square and triangular lattices, the UJ lattice hosts an additional critical phase between the FM and PM phases, which is stabilized by the combined effects of lattice inhomogeneity and geometric frustration.
This emergent critical phase may also be related to the fact that the AT model itself involves two coupled spin degrees of freedom, further enriching its phase structure.
This critical phase is characterized by quasi-long-range order, power-law decay of correlations, and BKT-type transitions at its boundaries.

\begin{table}[t]
  \centering
  \caption{Order parameter characteristics for different phases in the AT model on square and triangular  and UJ lattices. F: ferromagnetic, P: paramagnetic, AF: antiferromagnetic, C: critical.
  $p$AF: partial antiferromagnetic }
  \label{tab:phases_AT}
  \begin{tabular}{lcccc}
    \hline\hline
    ~~~Lattices ~& ~~Phases ~~& $~~~\langle\sigma\rangle$~~~ & $~~~\langle\tau\rangle$~~~ & $~~~\langle\sigma\tau\rangle~~~$ \\
\hline\hline
~~Square      & I     & F & F & F \\
          & II    & P & P & P \\
          & III   & P & P & F \\
          & IV    & P & P & AF \\
    \hline\hline
    ~~Triangular  & I     & F & F & F \\
      & II    & P & P & P \\
      & III   & P & P & F \\
    \hline\hline
      ~~UJ   & I     & F & F & F \\
      & II    & P & P & P \\
      & III   & P & P & $p$AF \\
      & IV    & \textbf{C} & \textbf{C} & \textbf{C} \\
      & V    & P & P & F \\
    \hline\hline
  \end{tabular}
\end{table}

\section{Discussion and conclusion}
\label{sec:con}

In this paper, the AT model on the UJ lattice is systematically studied, with a focus on its phase behavior and critical properties. The AT model on the UJ lattice is one of the few statistical models with discrete variables that exhibit a critical phase. Notably, for models with continuous spin variables (e.g., the XY model~\cite{XYmodel_UJ}, clock model~\cite{trg_UJ,tnclock8, tn-clock5, NN-clock6}, and other vortex excitation-related models), a non-zero-temperature quasi-long-range ordered phase has been observed on the UJ lattice in previous studies.However, for spin models with discrete variables (e.g., the Potts model), a quasi-long-range ordered phase can only emerge by artificially introducing the vortex energy~\cite{potts-vl4,potts7,Potts3} and other conditions such as \(J_1-J_2\) type of frustrations~\cite{Cheng_Chen,Cheng_Chen_1}.

Investigations in this paper reveal a novel critical phase characterized by a power-law decay of magnetization with system size, where the correlation length ratio $\xi/L$ remains finite even in the thermodynamic limit.
This emergent critical phase arises from the combined effects of lattice inhomogeneity, geometric frustration, and the two coupled spin degrees of freedom inherent to the AT model.

To corroborate the presence of a Berezinskii–Kosterlitz–Thouless  transition and determine the phase boundaries, the susceptibility $\widetilde{\chi} = \text{d}\langle m \rangle /\text{d}J$ is introduced as a key probe. Notably, pseudo-critical points $J_c(L)$ are observed to scale proportionally to $(\ln L)^{-2}$, a behavior consistent with Berezinskii–Kosterlitz–Thouless-type criticality. Furthermore, the critical exponent $\eta$ exhibits distinct values across the different sublattices of the coupled spin system.




Naturally,  several open questions remain. Given the high complexity of the phase diagram, characterized by numerous phase boundaries, we have not yet been able to determine the exact nature of many multiphase points. Furthermore, we have only discussed the phase diagram for \(J > 0\); the phase diagram for \(J < 0\) has not been presented here.  It also remains unclear whether the AT model applied to other inhomogeneous and frustrated lattices also exhibits a critical phase. Those questions will be discussed elsewhere.

In summary, our discovery of one of the few models with a critical phase may provide insights to understanding statistical physics models and potentially aid in the search for critical phases in Selenium adsorbed on
Ni surface~\cite{ex1} or Rydberg atom~\cite{atom1, atom2} experiments.
Since one-dimensional superfluids, superconductors, and supersolids all possess quasi-long-range order and fall into the category of critical phases. One-dimensional quantum spin systems are equivalent to two-dimensional classical spin systems and our results could also inspire the exploration of such quantum phases in one-dimensional quantum systems~\cite{Sachdev_2011}.


 {\it Acknowledgements:} The authors thank the anonymous referees for useful suggestions.
 The work was supported by the Hefei National Research Center for Physical Sciences at the Microscale (Grant No. KF2021002). W. Z. acknowledges the support of the Shanxi Province Science Foundation (Grant No. 202303021221029).  C. D. acknowledges support from the National Natural Science Foundation of China (NSFC) under Grant No. 11975024. Y. J. was supported by the National Natural Science Foundation of China (Grant No. 12275263) and the Natural Science Foundation of Fujian Province (Grant No. 2023J02032).\\
 \appendix
\section{Other details of the transitions}
\begin{figure}[h]
    \centering
\includegraphics[width=1\linewidth]{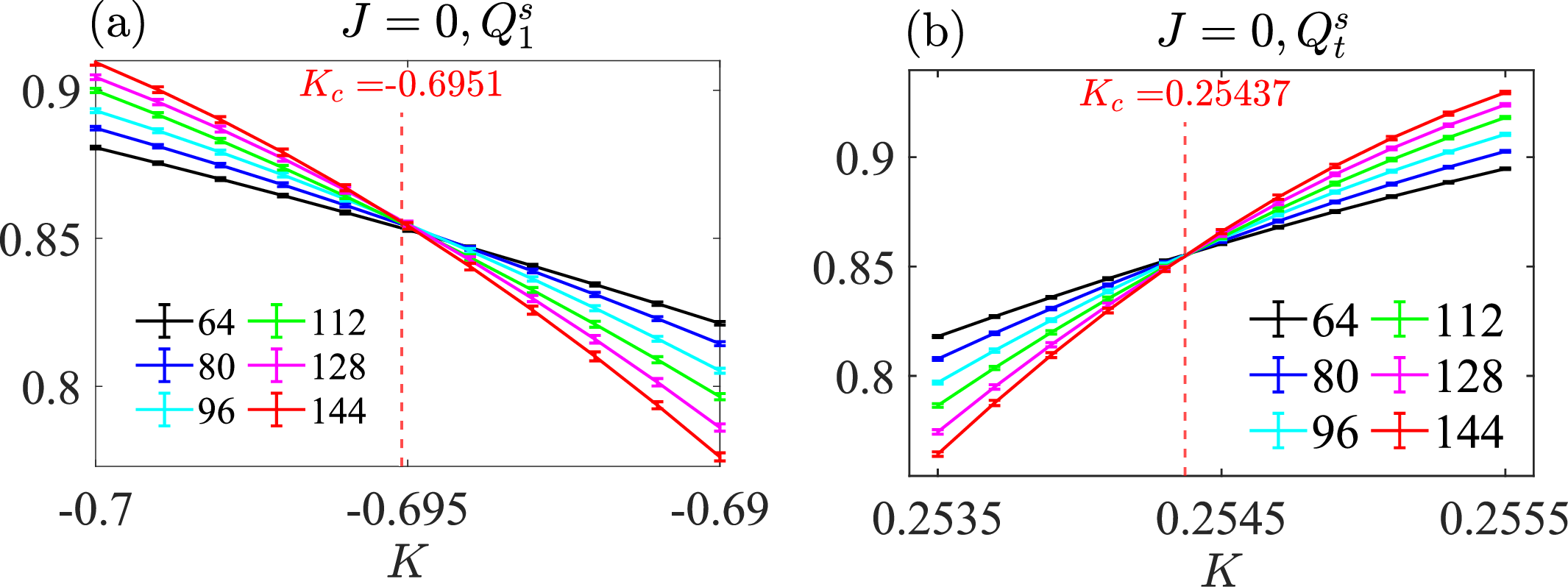}
  \caption{ Crossing behavior of the order parameter $Q_1$ at $J=0$ for various system sizes, indicating two phase transition points at $K_c = -0.695(1)$ (panel a) $y_t = 0.98(3)$ and $K_c= 0.254(1)$ (panel b) $y_t = 1.01(2)$.}
    \label{fig:J=0}
\end{figure}
The phase diagram we present contains numerous additional details; here we present the phase transitions occurring at \(J = 0\). 
This case is of particular interest due to the elegant closed-form expressions for the critical points~\cite{huang2011at}:
\begin{align}
K_c &= \frac{1}{2} \log\Bigg[ \frac{1}{\sqrt{2}} + \sqrt{\sqrt{2} - \frac{1}{2}} \Bigg] \nonumber \\
    &\approx 0.2543873426, \label{eq:Jc1} \\
K_c &= \frac{1}{2} \log\Bigg[ -\frac{1}{\sqrt{2}} + \sqrt{\sqrt{2} - \frac{1}{2}} \Bigg] \nonumber \\
    &\approx -0.6950741362. \label{eq:Jc2}
\end{align}
Figure~\ref{fig:J=0} shows the crossing of the order parameter \(Q_1\) at \(J=0\) for different system sizes, identifying two phase transition points at \(K_c =-0.695(1)\) (panel a) and \(K_c=0.254(1)\) (panel b). We perform finite-size scaling using Eq.~\ref{eq:q} with system sizes \(L = 64, 80, 96, 112, 128, 144\), and the crossing points are seen to converge to the critical values as the system size increases. The exact critical couplings are \(K_c = -0.69507\cdots\) and \(K_c = 0.25439\cdots\), which are in excellent agreement with the crossing points observed in the numerical data, confirming the locations of the phase transitions. The critical exponent $y_t$ further confirms that the transition belongs to the Ising universality class.

  \bibliography{references.bib}

@article{PhysRevB.49.8811,
  title = {Magnetization: A characteristic of the Kosterlitz-Thouless-Berezinskii transition},
  author = {Bramwell, S. T. and Holdsworth, P. C. W.},
  journal = {Phys. Rev. B},
  volume = {49},
  issue = {13},
  pages = {8811--8814},
  numpages = {0},
  year = {1994},
  month = {Apr},
  publisher = {American Physical Society},
  doi = {10.1103/PhysRevB.49.8811},
  url = {https://link.aps.org/doi/10.1103/PhysRevB.49.8811}
}

@article{sachdev1999quantum,
  title={Quantum phase transitions},
  author={Sachdev, Subir},
  journal={Physics world},
  volume={12},
  number={4},
  pages={33--38},
  year={1999}
}

@article{PhysRevB.66.024422,
  title = {Classical {H}eisenberg antiferromagnet away from the pyrochlore lattice limit: Entropic versus energetic selection},
  author = {Pinettes, C. and Canals, B. and Lacroix, C.},
  journal = {Phys. Rev. B},
  volume = {66},
  issue = {2},
  pages = {024422},
  numpages = {5},
  year = {2002},
  month = {Jul},
  publisher = {American Physical Society},
  doi = {10.1103/PhysRevB.66.024422},
  url = {https://link.aps.org/doi/10.1103/PhysRevB.66.024422}
}

@article{Bramwell_1993,
doi = {10.1088/0953-8984/5/4/004},
url = {https://doi.org/10.1088/0953-8984/5/4/004},
year = {1993},
month = {jan},
publisher = {},
volume = {5},
number = {4},
pages = {L53},
author = {S T Bramwell and P C W Holdsworth},
title = {Magnetization and universal sub-critical behaviour in two-dimensional XY magnets},
journal = {Journal of Physics: Condensed Matter}
}

@article{VALDES2007551,
title = {{±J Ising model on Dice lattices}},
journal = {Phys. A},
volume = {385},
number = {2},
pages = {551-557},
year = {2007},
issn = {0378-4371},
doi = {https://doi.org/10.1016/j.physa.2007.07.059},
url = {https://www.sciencedirect.com/science/article/pii/S0378437107008163},
author = {{J.F. Valdés and W. Lebrecht and E.E. Vogel}}
}

@book{Sachdev_2011, place={Cambridge}, edition={2}, title={Quantum Phase Transitions}, publisher={Cambridge University Press}, author={Sachdev, Subir}, year={2011}}

@article{zhu,
  title = {Gapless Coulomb State Emerging from a Self-Dual Topological Tensor-Network State},
  author = {Zhu, Guo-Yi and Zhang, Guang-Ming},
  journal = {Phys. Rev. Lett.},
  volume = {122},
  issue = {17},
  pages = {176401},
  numpages = {6},
  year = {2019},
  month = {Apr},
  publisher = {American Physical Society},
  doi = {10.1103/PhysRevLett.122.176401},
  url = {https://link.aps.org/doi/10.1103/PhysRevLett.122.176401}
}

@article{atom2,
  title = {Tunable Quantum Criticality in Multicomponent {R}ydberg Arrays},
  author = {Chepiga, Natalia},
  journal = {Phys. Rev. Lett.},
  volume = {132},
  issue = {7},
  pages = {076505},
  numpages = {5},
  year = {2024},
  month = {Feb},
  publisher = {American Physical Society},
  doi = {10.1103/PhysRevLett.132.076505},
  url = {https://link.aps.org/doi/10.1103/PhysRevLett.132.076505}
}

@Article{atom1,
author={Chepiga, Natalia
and Mila, Fr茅d茅ric},
title={{K}ibble-{Z}urek exponent and chiral transition of the period-4 phase of {R}ydberg chains},
journal={Nat. Commun.},
year={2021},
month={Jan},
day={18},
volume={12},
number={1},
pages={414},
issn={2041-1723},
doi={10.1038/s41467-020-20641-y},
url={https://doi.org/10.1038/s41467-020-20641-y}
}

@article{quantum_at,
  title = {Critical properties of the quantum {A}shkin-{T}eller chain with chiral perturbations},
  author = {L\"uscher, Bernhard E. and Mila, Fr\'ed\'eric and Chepiga, Natalia},
  journal = {Phys. Rev. B},
  volume = {108},
  issue = {18},
  pages = {184425},
  numpages = {7},
  year = {2023},
  month = {Nov},
  publisher = {American Physical Society},
  doi = {10.1103/PhysRevB.108.184425},
  url = {https://link.aps.org/doi/10.1103/PhysRevB.108.184425}
}

@article{percolation_at,
  title = {Geometric percolation of spins and spin dipoles in the {A}shkin-{T}eller model},
  author = {Banerjee, Aikya and Jana, Priyajit and Mohanty, P. K.},
  journal = {Phys. Rev. B},
  volume = {111},
  issue = {1},
  pages = {014403},
  numpages = {13},
  year = {2025},
  month = {Jan},
  publisher = {American Physical Society},
  doi = {10.1103/PhysRevB.111.014403},
  url = {https://link.aps.org/doi/10.1103/PhysRevB.111.014403}
}

@article{xy_3dat,
  title = {{X}{Y} criticality arising from emergent symmetry in the three-dimensional {A}shkin-{T}eller model},
  author = {Zhang, Dajun and Hu, Minghui and Sun, Yanan and Lv, Jian-Ping},
  journal = {Phys. Rev. E},
  volume = {111},
  issue = {4},
  pages = {044128},
  numpages = {7},
  year = {2025},
  month = {Apr},
  publisher = {American Physical Society},
  doi = {10.1103/PhysRevE.111.044128},
  url = {https://link.aps.org/doi/10.1103/PhysRevE.111.044128}
}

@article{spin_wave_UJ,
  title = {Quantum spin model with frustration on the {U}nion {J}ack lattice},
  author = {Collins, A. and McEvoy, J. and Robinson, D. and Hamer, C. J. and Weihong, Zheng},
  journal = {Phys. Rev. B},
  volume = {73},
  issue = {2},
  pages = {024407},
  numpages = {10},
  year = {2006},
  month = {Jan},
  publisher = {American Physical Society},
  doi = {10.1103/PhysRevB.73.024407},
  url = {https://link.aps.org/doi/10.1103/PhysRevB.73.024407}
}

@article{XYmodel_UJ,
  title = {Phase transitions in {X}{Y} antiferromagnets on plane triangulations},
  author = {Lv, Jian-Ping and Garoni, Timothy M. and Deng, Youjin},
  journal = {Phys. Rev. B},
  volume = {87},
  issue = {2},
  pages = {024108},
  numpages = {5},
  year = {2013},
  month = {Jan},
  publisher = {American Physical Society},
  doi = {10.1103/PhysRevB.87.024108},
  url = {https://link.aps.org/doi/10.1103/PhysRevB.87.024108}
}

@article{hsb_UJ,
  title = {Partially disordered {H}eisenberg antiferromagnet with short-range stripe correlations},
  author = {Blesio, G. G. and Lisandrini, F. T. and Gonzalez, M. G.},
  journal = {Phys. Rev. B},
  volume = {107},
  issue = {13},
  pages = {134418},
  numpages = {11},
  year = {2023},
  month = {Apr},
  publisher = {American Physical Society},
  doi = {10.1103/PhysRevB.107.134418},
  url = {https://link.aps.org/doi/10.1103/PhysRevB.107.134418}
}

@article{trg_UJ,
  title = {Tensor network renormalization approach to antiferromagnetic 6-state clock model on the {U}nion {J}ack lattice},
  author = {Homma, Kenji and Morita, Satoshi and Kawashima, Naoki},
  journal = {Phys. Rev. B},
  volume = {111},
  issue = {13},
  pages = {134427},
  numpages = {12},
  year = {2025},
  month = {Apr},
  publisher = {American Physical Society},
  doi = {10.1103/PhysRevB.111.134427},
  url = {https://link.aps.org/doi/10.1103/PhysRevB.111.134427}
}

@article{deng2011finite,
  title = {Finite-{T}emperature {P}hase {T}ransition in a {C}lass of {F}our-{S}tate {P}otts {A}ntiferromagnets},
  author = {Deng, Youjin and Huang, Yuan and Jacobsen, Jesper Lykke and Salas, Jes\'us and Sokal, Alan D.},
  journal = {Phys. Rev. Lett.},
  volume = {107},
  issue = {15},
  pages = {150601},
  numpages = {5},
  year = {2011},
  month = {Oct},
  publisher = {American Physical Society},
  doi = {10.1103/PhysRevLett.107.150601},
  url = {https://link.aps.org/doi/10.1103/PhysRevLett.107.150601}
}

@article{reduce,
  title = {Conformal Invariance of the Ising Model in Three Dimensions},
  author = {Deng, Youjin and Bl\"ote, Henk W. J.},
  journal = {Phys. Rev. Lett.},
  volume = {88},
  issue = {19},
  pages = {190602},
  numpages = {4},
  year = {2002},
  month = {Apr},
  publisher = {American Physical Society},
  doi = {10.1103/PhysRevLett.88.190602},
  url = {https://link.aps.org/doi/10.1103/PhysRevLett.88.190602}
}

@article{q_potts_xiang,
  title = {Partial Order and Finite-Temperature Phase Transitions in {P}otts Models on Irregular Lattices},
  author = {Chen, Q. N. and Qin, M. P. and Chen, J. and Wei, Z. C. and Zhao, H. H. and Normand, B. and Xiang, T.},
  journal = {Phys. Rev. Lett.},
  volume = {107},
  issue = {16},
  pages = {165701},
  numpages = {5},
  year = {2011},
  month = {Oct},
  publisher = {American Physical Society},
  doi = {10.1103/PhysRevLett.107.165701},
  url = {https://link.aps.org/doi/10.1103/PhysRevLett.107.165701}
}

@article{dingUJ,
  title = {Critical properties of the {H}intermann-{M}erlini model},
  author = {Ding, Chengxiang and Wang, Yancheng and Zhang, Wanzhou and Guo, Wenan},
  journal = {Phys. Rev. E},
  volume = {88},
  issue = {4},
  pages = {042117},
  numpages = {7},
  year = {2013},
  month = {Oct},
  publisher = {American Physical Society},
  doi = {10.1103/PhysRevE.88.042117},
  url = {https://link.aps.org/doi/10.1103/PhysRevE.88.042117}
}

@article{worm,
  title = {Worm-type {M}onte {C}arlo simulation of the {A}shkin-{T}eller model on the triangular lattice},
  author = {Lv, Jian-Ping and Deng, Youjin and Chen, Qing-Hu},
  journal = {Phys. Rev. E},
  volume = {84},
  issue = {2},
  pages = {021125},
  numpages = {11},
  year = {2011},
  month = {Aug},
  publisher = {American Physical Society},
  doi = {10.1103/PhysRevE.84.021125},
  url = {https://link.aps.org/doi/10.1103/PhysRevE.84.021125}
}

@article{ex1,
  title = {Phase Diagram of {Selenium} Adsorbed on the {Ni(100)} Surface: A Physical Realization of the {A}shkin-{T}eller Model},
  author = {Bak, Per and Kleban, P. and Unertl, W. N. and Ochab, J. and Akinci, G. and Bartelt, N. C. and Einstein, T. L.},
  journal = {Phys. Rev. Lett.},
  volume = {54},
  issue = {14},
  pages = {1539--1542},
  year = {1985},
  month = {Apr},
  publisher = {American Physical Society},
  doi = {10.1103/PhysRevLett.54.1539},
  url = {https://link.aps.org/doi/10.1103/PhysRevLett.54.1539}
}

@article{26,
  title={Phase transition in the three-state {P}otts antiferromagnet on the diced lattice},
  author={Koteck{\`y}, Roman and Salas, Jes{\'u}s and Sokal, Alan D},
  journal={Phys. Rev. Lett.},
  volume={101},
  number={3},
  pages={030601},
  year={2008},
  publisher={APS},
doi = {10.1103/PhysRevLett.101.030601},
  url = {https://link.aps.org/doi/10.1103/PhysRevLett.101.030601}
}

@article{32,
  title={A new Kempe invariant and the (non)-ergodicity of the Wang--Swendsen--Koteck{\`y} algorithm},
  author={Mohar, Bojan and Salas, Jes{\'u}s},
  journal={Journal of Physics A: Mathematical and Theoretical},
  volume={42},
  number={22},
  pages={225204},
  year={2009},
  publisher={IOP Publishing},
 url={http://dx.doi.org/10.1088/1751-8113/42/22/225204},
   DOI={10.1088/1751-8113/42/22/225204}
}

@article{24,
doi = {10.1088/0305-4470/8/9/020},
url = {https://dx.doi.org/10.1088/0305-4470/8/9/020},
year = {1975},
month = {sep},
publisher = {},
volume = {8},
number = {9},
pages = {1508},
author = {H J F Knops},
title = {A branch point in the critical surface of the {A}shkin-{T}eller model in the renormalization group theory},
journal = {J. Phys. A: Math. Gen.}
}

@book{4, place={Cambridge}, edition={3}, title={A Guide to Monte Carlo Simulations in Statistical Physics}, publisher={Cambridge University Press}, author={Landau, David P. and Binder, Kurt}, year={2009},
doi={
https://doi.org/10.1017/CBO9781139696463}}

@article{8,
doi = {10.1088/0305-4470/22/10/018},
url = {https://dx.doi.org/10.1088/0305-4470/22/10/018},
year = {1989},
month = {may},
publisher = {},
volume = {22},
number = {10},
pages = {1639},
author = {J Chahine and  J R Drugowich de Felicio and  N Caticha},
title = {Non-universal exponents and marginal operators via a Monte Carlo renormalisation group},
journal = {Journal of Physics A: Mathematical and General}
}

@article{36,
  title={Dynamic critical behavior of a {S}wendsen-{W}ang-type algorithm for the Ashkin-Teller model},
  author={Salas, Jes{\'u}s and Sokal, Alan D},
  journal={J. Stat. Phys.},
  volume={85},
  pages={297--361},
  year={1996},
  publisher={Springer},
  doi = {https://doi.org/10.1007/BF02174209},
  url = {https://link.springer.com/article/10.1007/BF02174209}
}

@article{23,
  title = {Critical {I}sing lines of the $d=2$ {A}shkin-{T}eller model},
  author = {Kamieniarz, G. and Koz\l{}owski, P. and Dekeyser, R.},
  journal = {Phys. Rev. E},
  volume = {55},
  issue = {3},
  pages = {3724--3726},
  numpages = {0},
  year = {1997},
  month = {Mar},
  publisher = {American Physical Society},
  doi = {10.1103/PhysRevE.55.3724},
  url = {https://link.aps.org/doi/10.1103/PhysRevE.55.3724}
}

@article{9,
  title = {Phase diagram for the {A}shkin-{T}eller model in three dimensions},
  author = {Ditzian, Ruth V. and Banavar, Jayanth R. and Grest, G. S. and Kadanoff, Leo P.},
  journal = {Phys. Rev. B},
  volume = {22},
  issue = {5},
  pages = {2542--2553},
  numpages = {0},
  year = {1980},
  month = {Sep},
  publisher = {American Physical Society},
  doi = {10.1103/PhysRevB.22.2542},
  url = {https://link.aps.org/doi/10.1103/PhysRevB.22.2542}
}

@article{ashkin1943,
  title = {Statistics of Two-Dimensional Lattices with Four Components},
  author = {Ashkin, J. and Teller, E.},
  journal = {Phys. Rev.},
  volume = {64},
  issue = {5-6},
  pages = {178--184},
  numpages = {0},
  year = {1943},
  month = {Sep},
  publisher = {American Physical Society},
  doi = {10.1103/PhysRev.64.178},
  url = {https://link.aps.org/doi/10.1103/PhysRev.64.178}
}

@article{PRB,
  title = {Monte Carlo simulation with tensor network states},
  author = {Wang, Ling and Pi\ifmmode \check{z}\else \v{z}\fi{}orn, Iztok and Verstraete, Frank},
  journal = {Phys. Rev. B},
  volume = {83},
  issue = {13},
  pages = {134421},
  numpages = {6},
  year = {2011},
  month = {Apr},
  publisher = {American Physical Society},
  doi = {10.1103/PhysRevB.83.134421},
  url = {https://link.aps.org/doi/10.1103/PhysRevB.83.134421}
}

@article{Metropolis1953,
  author = {Metropolis, Nicholas and Rosenbluth, Arianna W. and Rosenbluth, Marshall N. and Teller, Augusta H. and Teller, Edward},
  title = {Equation of State Calculations by Fast Computing Machines},
 journal = {J. Chem. Phys.},
  volume = {21},
  number = {6},
  pages = {1087--1092},
  year = {1953},
  doi = {10.1063/1.1699114}
}

@inproceedings{bm,
  author    = {Broadbent, Simon R and Hammersley, John M},
  booktitle = {Math. Proc. Camb. Philos. Soc.},
  isbn      = {1469-8064},
  number    = {3},
  pages     = {629--641},
  publisher = {Cambridge University Press},
  title     = {{Percolation processes: I. Crystals and mazes}},
  volume    = {53},
  year      = {1957},
  url       = {https://www.semanticscholar.org/paper/Percolation-processes.-I.-Crystals-and-Mazes-Broadbent-Hammersley/55caf92bd5d58d6c2f5593ceccb2fd1916fd5340}
}

@article{Potts3,
  title = {Overlap of two topological phases in the antiferromagnetic {P}otts model},
  author = {Zhao, Ran and Ding, Chengxiang and Deng, Youjin},
  journal = {Phys. Rev. E},
  volume = {97},
  issue = {5},
  pages = {052131},
  numpages = {5},
  year = {2018},
  month = {May},
  publisher = {American Physical Society},
  doi = {10.1103/PhysRevE.97.052131},
  url = {https://link.aps.org/doi/10.1103/PhysRevE.97.052131}
}

@article{potts-vl4,
  title = {Quasi-Long-Range Order and Vortex Lattice in the Three-State {Potts} Model},
  author = {Bhattacharya, Soumyadeep and Ray, Purusattam},
  journal = {Phys. Rev. Lett.},
  volume = {116},
  issue = {9},
  pages = {097206},
  numpages = {5},
  year = {2016},
  month = {Mar},
  publisher = {American Physical Society},
  doi = {10.1103/PhysRevLett.116.097206},
  url = {https://link.aps.org/doi/10.1103/PhysRevLett.116.097206}
}

@article{tn-clock5,
  title = {Critical properties of the two-dimensional $q$-state clock model},
  author = {Li, Zi-Qian and Yang, Li-Ping and Xie, Z. Y. and Tu, Hong-Hao and Liao, Hai-Jun and Xiang, T.},
  journal = {Phys. Rev. E},
  volume = {101},
  issue = {6},
  pages = {060105},
  numpages = {5},
  year = {2020},
  month = {Jun},
  publisher = {American Physical Society},
  doi = {10.1103/PhysRevE.101.060105},
  url = {https://link.aps.org/doi/10.1103/PhysRevE.101.060105}
}

@article{NN-clock6,
doi = {10.1088/1367-2630/ac63da},
url = {https://dx.doi.org/10.1088/1367-2630/ac63da},
journal = {N. J. Phys.}, 
year = {2022},
month = {apr},
publisher = {IOP Publishing},
volume = {24},
number = {4},
pages = {043040},
author = {Giataganas, Dimitrios and Huang, Ching-Yu and Lin, Feng-Li},
title = {Neural network flows of low $q$-state {P}otts and clock models},
journal = {New Journal of Physics}
}

@article{potts7,
  title = {Reentrance of {B}erezinskii-{K}osterlitz-{T}houless-like transitions in a three-state {P}otts antiferromagnetic thin film},
  author = {Ding, Chengxiang and Guo, Wenan and Deng, Youjin},
  journal = {Phys. Rev. B},
  volume = {90},
  issue = {13},
  pages = {134420},
  numpages = {6},
  year = {2014},
  month = {Oct},
  publisher = {American Physical Society},
  doi = {10.1103/PhysRevB.90.134420},
  url = {https://link.aps.org/doi/10.1103/PhysRevB.90.134420}
}

@article{tnclock8,
  title = {Tensor-network renormalization approach to the $q$-state clock model},
  author = {Li, Guanrong and Pai, Kwok Ho and Gu, Zheng-Cheng},
  journal = {Phys. Rev. Res.},
  volume = {4},
  issue = {2},
  pages = {023159},
  numpages = {19},
  year = {2022},
  month = {May},
  publisher = {American Physical Society},
  doi = {10.1103/PhysRevResearch.4.023159},
  url = {https://link.aps.org/doi/10.1103/PhysRevResearch.4.023159}
}

@article{k,
    doi = {10.1088/0031-9112/34/4/045},
    url = {https://dx.doi.org/10.1088/0031-9112/34/4/045},
    year = {1983},
    month = apr,
    volume = {34},
    number = {4},
    pages = {167},
    author = {Moore, M A},
    title = {Exactly Solved Models in Statistical Mechanics},
    journal = {Physics Bulletin}
}

@article{log,
  title = {Extraordinary-log Universality of Critical Phenomena in Plane Defects},
  author = {Sun, Yanan and Hu, Minghui and Deng, Youjin and Lv, Jian-Ping},
  journal = {Phys. Rev. Lett.},
  volume = {131},
  issue = {20},
  pages = {207101},
  numpages = {6},
  year = {2023},
  month = {Nov},
  publisher = {American Physical Society},
  doi = {10.1103/PhysRevLett.131.207101},
  url = {https://link.aps.org/doi/10.1103/PhysRevLett.131.207101}
}

@bachelorsthesis{huang2011at,
    author = {Huang, Yuan},
    title = {Phase Diagram of the {AT} Model on the {U}nion-{J}ack Lattice, {B}achelor's {T}hesis},
    school = {University of Science and Technology of China},
    year = {2011},
    month = {June},
    note = {{S}upervisor: Professor Youjin Deng},
    address = {Hefei}
}

@article{17_1,
  title = {Monte Carlo studies of chiral and spin ordering of the three-dimensional {Heisenberg} spin glass},
  author = {Viet, Dao Xuan and Kawamura, Hikaru},
  journal = {Phys. Rev. B},
  volume = {80},
  issue = {6},
  pages = {064418},
  numpages = {20},
  year = {2009},
  month = {Aug},
  publisher = {American Physical Society},
  doi = {10.1103/PhysRevB.80.064418},
  url = {https://link.aps.org/doi/10.1103/PhysRevB.80.064418}
}

@article{20_1,
  title = {Critical behavior of the three-dimensional {Ising} spin glass},
  author = {Ballesteros, H. G. and Cruz, A. and Fern\'andez, L. A. and Mart\'{\i}n-Mayor, V. and Pech, J. and Ruiz-Lorenzo, J. J. and Taranc\'on, A. and T\'ellez, P. and Ullod, C. L. and Ungil, C.},
  journal = {Phys. Rev. B},
  volume = {62},
  issue = {21},
  pages = {14237--14245},
  numpages = {0},
  year = {2000},
  month = {Dec},
  publisher = {American Physical Society},
  doi = {10.1103/PhysRevB.62.14237},
  url = {https://link.aps.org/doi/10.1103/PhysRevB.62.14237}
}

@article{21_1,
doi = {10.1088/1742-5468/2008/08/P08003},
url = {https://dx.doi.org/10.1088/1742-5468/2008/08/P08003},
year = {2008},
month = {aug},
publisher = {},
volume = {2008},
number = {08},
pages = {P08003},
journal = {J. Stat. Mech.}, 
author = {Hasenbusch, Martin},
title = {The Binder cumulant at the {K}osterlitz–{T}houless transition},
journal = {Journal of Statistical Mechanics: Theory and Experiment}
}

@article{22_1,
  title = {Correlation length in a generalized two-dimensional {XY model}},
  author = {Nui, Duong Xuan and Tuan, Le and Trung Kien, Nguyen Duc and Huy, Pham Thanh and Dang, Hung T. and Viet, Dao Xuan},
  journal = {Phys. Rev. B},
  volume = {98},
  issue = {14},
  pages = {144421},
  numpages = {9},
  year = {2018},
  month = {Oct},
  publisher = {American Physical Society},
  doi = {10.1103/PhysRevB.98.144421},
  url = {https://link.aps.org/doi/10.1103/PhysRevB.98.144421}
}

@article{Cheng_Chen,
  title = {{B}erezinskii-{K}osterlitz-{T}houless phase transitions in a kagome spin ice by a quantifying {M}onte {C}arlo process: Distribution of Hamming distances},
  author = {Su, Wen-Yu and Hu, Feng and Cheng, Chen and Ma, Nvsen},
  journal = {Phys. Rev. B},
  volume = {108},
  issue = {13},
  pages = {134422},
  numpages = {13},
  year = {2023},
  month = {Oct},
  publisher = {American Physical Society},
  doi = {10.1103/PhysRevB.108.134422},
  url = {https://link.aps.org/doi/10.1103/PhysRevB.108.134422}
}

@article{Weigel_2005_F_model,
   title={The square-latticeFmodel revisited: a loop-cluster update scaling study},
   volume={38},
   ISSN={1361-6447},
   url={http://dx.doi.org/10.1088/0305-4470/38/32/002},
   DOI={10.1088/0305-4470/38/32/002},
   number={32},
   journal={J. Phys. A: Math. Gen.},
   publisher={IOP Publishing},
   author={Weigel, M and Janke, W},
   year={2005},
   month=jul, pages={7067–7092} }

@article{ChenTao.111.094201,
  title = {{T}ensor network {M}onte {C}arlo simulations for the two-dimensional random-bond {I}sing model},
  author = {Chen, Tao and Guo, Erdong and Zhang, Wanzhou and Zhang, Pan and Deng, Youjin},
  journal = {Phys. Rev. B},
  volume = {111},
  issue = {9},
  pages = {094201},
  numpages = {14},
  year = {2025},
  month = {Mar},
  publisher = {American Physical Society},
  doi = {10.1103/PhysRevB.111.094201},
  url = {https://link.aps.org/doi/10.1103/PhysRevB.111.094201}
}

@article{7Parisen_Toldin_2009,
   title={Strong-Disorder Paramagnetic-Ferromagnetic Fixed Point in the Square-Lattice ±{$J$} {I}sing {M}odel},
   volume={135},
   ISSN={1572-9613},
   url={http://dx.doi.org/10.1007/s10955-009-9705-5},
   DOI={10.1007/s10955-009-9705-5},
   number={5–6},
   journal={J. Stat. Phys.},
   publisher={Springer Science and Business Media LLC},
   author={Parisen Toldin, Francesco and Pelissetto, Andrea and Vicari, Ettore},
   year={2009},
   month=mar, pages={1039–1061} }

@book{18,
  title={Tilings and patterns},
  author={Gr{\"u}nbaum, Branko and Shephard, Geoffrey Colin},
  year={1987},
  publisher={Courier Dover Publications},
  url = {https://simetri-graphics.github.io/simetri/books/tilings_and_patterns/}
}

@article{Cheng_Chen_1,
    author = {Chern, Gia-Wei and Tchernyshyov, Oleg},
    title = {Magnetic charge and ordering in kagome spin ice},
    journal = {Phil. Trans. R. Soc. A},
    volume = {370},
    number = {1981},
    pages = {5718-5737},
    year = {2012},
    month = {12},
    issn = {1364-503X},
    doi = {10.1098/rsta.2011.0388},
    url = {https://doi.org/10.1098/rsta.2011.0388},
}

\end{document}